%% file: template-arxiv.tex
\documentclass[a4paper,11pt]{article}
\usepackage{amsmath, amsfonts, amssymb, amsthm}
\usepackage{graphicx}
\usepackage{fancyhdr}
\usepackage{hyperref}
\usepackage{float}
\usepackage{url}
\usepackage[utf8]{inputenc}
\usepackage[top=0.75in, bottom=1in, left=0.75in, right=0.75in]{geometry}
\usepackage{bbm} 
\usepackage{xcolor}
\usepackage{adjustbox}
\usepackage{stackrel}
\usepackage{csquotes}
\usepackage{caption}
\usepackage[american]{babel}
\usepackage[bibstyle=apa,citestyle=numeric,natbib=true]{biblatex}
\DeclareLanguageMapping{american}{american-apa}
\DeclareFieldFormat{labelnumberwidth}{[#1]}
\allowdisplaybreaks
\defbibenvironment{bibliography}
  {\list 
    {\printfield[labelnumberwidth]{labelnumber}} 
  {\setlength{\labelwidth}{\labelnumberwidth}%
   \setlength{\leftmargin}{0pt}%
   \setlength{\labelsep}{\biblabelsep}%
   \setlength{\itemsep}{\bibitemsep}%
   \setlength{\itemindent}{\labelwidth}%
   \addtolength{\itemindent}{\labelsep}%
   \setlength{\parsep}{\bibparsep}}%
   } 
{\endlist} 
{\item}
\input{commands}

\providecommand{\keywords}[1]
{
  \small	
  \textbf{\textit{Keywords---}} #1
}

\title{Optimal portfolios with anticipating information on the stochastic interest rate}
\author{Bernardo D'Auria
\thanks{Department of Mathematics “Tullio Levi-Civita”, University of Padova, 35131, Padova PD, Italy.\\
\email{bernardo.dauria@unipd.it}.}
\and Jos\'{e} A. Salmer\'{o}n
\thanks{Directorate General for the Regulation of Gambling, 
Ministry of Social Rights, Consumer Affaires and 2030 Agenda, 28012, Madrid, Spain. 
 \email{jasalmeron@uordenacionjuego.gob.es}.}
}

\numberwithin{equation}{section}
\usepackage{color}
\usepackage{amsmath,amsfonts,mathtools}
\usepackage{comment}
\usepackage{multirow} 
\usepackage{enumitem} 
\usepackage{standalone} 
\usepackage{etoolbox} 
\usepackage{xspace} 
\newtheorem{theorem}{Theorem}[section]
\newtheorem{example}[theorem]{Example}
\newtheorem{lemma}[theorem]{Lemma}
\newtheorem{proposition}[theorem]{Proposition}
\newtheorem{corollary}[theorem]{Corollary}
\newtheorem{definition}[theorem]{Definition}
\newtheorem{assumption}[theorem]{Assumption}

\newtheorem{remark}[theorem]{Remark}

\providecommand{\citep}[1]{(~\cite{#1})}
\newcommand{\email}[1]{\href{mailto:#1}{#1}}

\begin{document}

\maketitle
\begin{abstract}
By employing the technique of enlargement of filtrations, we demonstrate how to incorporate information about the future trend of the stochastic interest rate process into a financial model. By modeling the interest rate as an affine diffusion process, we obtain explicit formulas for the additional expected logarithmic utility in solving the optimal portfolio problem.

We begin by solving the problem when the additional information directly refers to the interest rate process, and then extend the analysis to the case where the information relates to the values of an underlying Markov chain. The dynamics of this chain may depend on anticipated market information, jump at predefined epochs, and modulate the parameters of the stochastic interest rate process. The theoretical study is then complemented by an illustrative numerical analysis.\end{abstract}

\keywords{
{Stochastic control};
{Portfolio Choice};
Enlargement of filtrations;
\Vmod{} short-rate model; 
Additional logarithmic expected utility;
Markovian modulation.
}

\maketitle
\section{Introduction}\label{sec:intro}
We address the optimal portfolio problem in a market in which a trading agent has some anticipating information on the short rate process.
In the literature, it is more common to find examples in which the anticipating information considers the future trend of some risky assets.
We refer to~\cite{Filipovic09} and~\cite{Privault2012interestrate} for a detailed introduction on stochastic interest rate models.
The availability of anticipating information creates an asymmetry in the market and to analyze it, we assume that there are two agents and only one of them is allowed to make use of the additional information.

The extension of the optimal portfolio problem to the case of anticipating information is generally dealt with by the technique of 
\emph{enlargement of filtrations} that consists in analyzing the problem in a new filtration obtained by extending the natural one with the additional information. 
In order to ensure that in the new filtration all former martingales remain in the class of semimartingales, 
we assume that the random variable modeling the new information satisfies the so-called \emph{Jacod's hypothesis}, see Assumption~\ref{ass.jacod} below.

We focus on the \emph{initial enlargements}, where the additional information is revealed to the informed agent from the beginning. 
This procedure was originally applied to the optimal portfolio problem in~\cite{karatzas1996}.

Assuming that the anticipating information was modeled by a random variable, say $G$, satisfying Jacod's hypothesis, they explicitly solved for the optimal investment strategy in a variety of examples.
\cite{AmendingerImkellerSchweizer1998} showed that the additional expected logarithmic utility is equal to the entropy of $G$ when it is represented by an atomic random variable.
Moreover, whenever this random variable is not purely atomic, 
the additional gain is shown to be unbounded.
\cite{AmendingerImkellerSchweizer1998} introduced the information drift, below denoted by~$\alpha^G$, to express the semimartingale representation of the assets in the enlarged filtration, and its norm in~$L^2(dt\times\PP)$ was linked with the additional gains measured with the logarithmic utility. 
\cite{Amendinger00} considers the conditions that guarantee a martingale representation in the enlarged filtrations, in particular assuming the validity of Jacod's hypothesis in a closed time interval.
Then, in~\cite{AnkirchnerImkeller06}, an explicit computation of the information drift is given in a very general situation.

A different approach was taken in~\cite{Biagini2005}.
By assuming the existence of a local
maximal solution, a semimartingale decomposition of the Brownian motion
driving the risky asset is provided in the enlarged filtration.

We mention~\cite{JAPoksendal_2007}, in which the authors introduce additional constraints to the set of admissible strategies in order to bound the additional expected logarithmic utility in presence of insider information.
In~\cite{Amendinger2003}, the optimal portfolio problem is solved for utility functions different from the logarithmic one (CRRA and exponential)
and the price of the information is introduced as the fair price that an investor would initially pay to acquire the additional information.
In~\cite{IMKELLER2000}, Jacod's hypothesis is weakened by solving the optimal portfolio problem via the Malliavin calculus, while in~\cite{AnkirchnerImkeller05}, they discuss the relationship between having additional information and the financial opportunities of arbitrage.
Computing the value of the additional information is an active topic of research, as shown by recent results such as in~\cite{Fontana18}, where the authors quantify 
the value of arbitrage opportunities in a general semimartingale context, 
or as in~\cite{DauriaSalmeron2020}, where they analyze the relationships between arbitrage, 
initial enlargements and additional expected logarithmic utility. 
In~\cite{Rogers2020}, the authors compute the value of the information when this refers to the jumps of a Poisson process, and
in~\cite{DAuriaSalmeronANOR}, the additional expected logarithmic gains are obtained  in a mixed Brownian-Poisson market with a general set-up.
Recently, in~\cite{Halconruy23DEF}, the author considers the additional information in a discrete-time market.

In this paper, even if technically we employ techniques similar to the ones developed in the references given above, 
we use specialized ones to deal with the specific case in which the interest rate is stochastic and not adapted to the filtration generated by the price processes, and when the additional information, 
which is revealed to the agent, is about characteristics of its future trend.
Our aim is to compute the optimal strategy in presence of anticipating information and to give the expected logarithmic gain that an agent could get when the information refers to the future dynamics of the interest rate.
An example of an optimal portfolio problem under the stochastic interest rate -- but without additional information -- is considered in~\cite{JAPdeelstra_grasselli_koehl_2000},
see also~\cite{Zhang23DEF}, where they employ a BSDE approach.

We analyze two different models dealing with anticipating information on the stochastic interest rate. 
In the first model, we define the interest rate as a general affine diffusion process, 
as this structure is quite flexible and allows us to get quite explicit expressions. 
Moreover this class includes, as a special case, 
the well-known \Vmod{} model, introduced in~\cite{vasicek}. 
This model allows us to simplify the computations involving the distribution of the process.

In the second model, we introduce an anticipating discrete time Markov chain that modulates the evolution of the interest rate. 
The existence of such modulation in the market has been deeply studied, 
however we believe that this is the first time that this modulation has an anticipating nature.
This model is sometimes referred to as a regime-switching model, because it leads to scenarios in which the market coefficients switch from some values (e.g.\ bull market) 
to other ones (e.g.\ bear market), to then switch back.
A seminal paper solving the modulated version of the optimal portfolio problem is~\cite{Bauerle}.
Explicit solutions for the investment/consumption problem with HARA utilities were obtained in~\cite{SotomayorCadenillas2009}.
A similar model, where the modulation is generated by a semi-Markov process, was analyzed in~\cite{GhoshGoswamiKumar2009}.
In~\cite{Azevedo2014}, a dynamic programming principle was derived for a 
Markov-switching jump–diffusion process. 
In~\cite{AntonelliRamponiScarlattijournal}, they studied the VaR on zero-coupon bonds for a regime-switching model, while in~\cite{Ferrari18} the regime switching is motivated by an irreversible extraction problem.

With respect to the first model, the main  contribution of this work is in focusing on problems where the anticipating information is about the short rate, by providing explicit formulas and solving some examples with full details.
Moreover, in Section~\ref{subsec:non.atomic}, we show how to apply, for some specific type of information, a technique based on vector measure theory.
As for the second model, our main contribution is in introducing, as far as we know for the first time, a Markov modulated model whose dynamics may be anticipating.
To show a non-trivial example of this new model, we consider the case in which the anticipating information is given by the distance of the Brownian motion with respect to its running minimum, see Example~\ref{Ex.min}.

The paper is organized as follows.
In Section~\ref{sec:model.notation}, we introduce in detail the general model where the interest rate process is represented by an affine diffusion.
In Section~\ref{sec:opt.prob}, we provide the semimartingale decomposition of the risky asset in the enlarged filtration, and we solve the optimal portfolio problem under the logarithmic utility.
In Section~\ref{sec:non.lin.info}, we introduce a methodology for  explicitly computing the information drift in a more general setting by employing vector measures.
We use vector measures to represent purely atomic random variables, for instance when the additional information consists in knowing  an upper (or lower) bound for the  value of the interest rate at a final horizon time. 
To show an example in which the information is given by a continuous random variable, we consider an initial enlargement of filtrations that includes at the initial time, the value of the $n$-th power of the terminal value of the Brownian motion driving the interest rate process.
In Section~\ref{sec:markov.modul}, we introduce a Markov chain which modulates the evolution of the interest rate process by modifying the parameters of its affine diffusion representation.
We provide the additional expected logarithmic gains in this set-up.
In Section~\ref{sec:num.illustr}, we present some numerical examples and finally, in Section~\ref{sec:conclusions}, we end with some concluding remarks.

\section{Affine model and notation}\label{sec:model.notation}
As a general set-up, we consider an underlying probability space
$(\Omega, \cF,\bF,\PP)$,
where~$\cF$ is the $\sigma$-algebra 
and $\bF=\{\cF_t, t\geq0\}$ is the augmentation of the natural filtration of the process $(B^R,B^S)=((B^R_t,B^S_t), t\geq0)$, which is a bi-dimensional process whose components are $\bF$-Brownian motions with constant correlation $\rho\neq 0$.
Sometimes we use the notation  $\bF^R$ and $\bF^S$
(resp. $\cF_T^R$ and $\cF_T^S$) to denote the filtrations (resp. $\sigma$-algebras) individually generated by the processes $B^R$ and $B^S$.
We fix a finite horizon time $T>0$, that is,  the trader is allowed to invest in the time interval $[0,T]$.

To make the analysis simpler, we consider a portfolio made of only two assets, $S=\prT{S}$ and $D=\prT{D}$.
Both processes are $\bF$-semimartingales, and their dynamics are defined by the following SDEs,
\begin{subequations}\label{DS.SDE}
\begin{align}
dD_t &= D_t \, R_t \, dt,\quad  D_0 = 1
\label{D.SDE}\\
dS_t &= S_t \, \left(\eta_t dt + \xi_t dB^S_t\right),\ \,\quad  S_0 = s_0 > 0  \label{S.SDE}
\end{align}
\end{subequations}
where the stochastic process $R=\prT{R}$ is the \emph{instantaneous interest rate}, also referred to as the \emph{short rate},
see Chapter 5 of~\cite{Filipovic09} or~\cite{GibsonLhabitantTalay2010}.
The proportional drift, $\eta=\prT{\eta}$, and the volatility, $\xi=\prT{\xi}$, of $S$ are assumed to be $\bF$-predictable processes satisfying the following condition
\begin{equation*}
    \EE\left[\int_0^T\left(\frac{\eta_t - R_t}{\xi_t}\right)^2 dt\right]<+\infty.
\end{equation*}
In addition, we assume that the condition of  
\emph{No Free Lunch Vanishing Risk~(NFLVR)} 
is true under the filtration~$\bF$.

The interest rate process $R$ is assumed to be an \emph{affine diffusion}, satisfying the following SDE,
\begin{align}\label{R.SDE}
dR_t =  \left[ a_1(t)R_t+a_2(t) \right]dt + b_2(t)dB^R_t ,\quad R_0 = r_0 > 0,
\end{align}
where the functions $a_1, a_2, b_2$ are supposed to be deterministic and bounded. 
The benefit of working with this class is that it allows for a quite explicit formula for the conditional distribution of $(R_t\vert R_s)$, see Lemma~\ref{lema2} below.
Moreover, the \OU{} process, introduced as a financial model for the short rate in~\cite{vasicek}, belongs to this class. 
See also~\cite{Ferrari17}, in which the process models exchange rates.
For convenience, we denote the \OU{} process by $Y=\prT{Y}$, 
which satisfies the following well-known~SDE
 \begin{align}\label{Y.SDE}
 dY_t = k(\mu-Y_t)dt +\sigma dB^Y_t ,\quad Y_0 = y_0 > 0,
 \end{align}
 with $\mu\in\bR$ and $k>0$, $\sigma>0$ given constants.

The class of processes represented by~\eqref{R.SDE} is slightly different from the affine diffusion class defined in~\cite{Alfonsi2015} and in~\cite{Filipovic09}, as more generally,
we allow for structural terms that are non-homogeneous in time, 
while less generally, we do not allow the diffusion term to depend linearly on the value of the process $R$.

Under the above hypothesis, we assume that an investor employs a strategy $\pi=\prT{\pi}$
with the aim of optimizing the expected logarithmic utility of the wealth at a finite terminal time $T>0$.
The strategy $\pi$ makes optimal use of all information at disposal of the agent at each instant, and we assume it is adapted to a filtration 
$\bH=\{\cH_t, t\geq0\}$ possibly larger than filtration $\bF$, that is  $\bH \supseteq \bF$.
We next define the set of all admissible strategies, $\cA(\bH)$, among which the trader can choose the one that best fits her objective function.

\begin{definition}
The set of admissible strategies $\cA(\bH)$ is composed of all the self-financing $\bH$-adapted portfolios $\pi$ such that
$\int_0^T  \xi_t^2\pi_t^2 dt < +\infty$, $\PP$-a.s.
\end{definition}

Denoting by $X^\pi=\prT{X^\pi}$ the wealth of the portfolio of the investor under an admissible strategy $\pi$,
we can write its dynamics as the solution of the following SDE
\begin{align*}
\frac{dX_t^{\pi}}{X_t^{\pi}} &= (1-\pi_t)\frac{dD_t}{D_t} + \pi_t \frac{dS_t}{S_t},
\quad X_0=x_0 > 0 ,
\end{align*}
and by~\eqref{DS.SDE}, it can be expressed in the following form
\begin{align}\label{X.SDE}
dX_t^{\pi} &= (1-\pi_t)X_t^{\pi}R_t \, dt + \pi_t X_t^{\pi}\left(\eta_t \, dt + \xi_t \, dB^S_t\right) ,
\quad X_0 = x_0.
\end{align}
We define $\bV_t^{\bH}$, as the optimal expected value of the portfolio at time $t\in(0,T)$, given the information flow $\bH$, that is  
\begin{align}\label{utility}
\bV_t^{\bH} \defeq \sup_{\pi\in\cA({\bH})} \EE[\ln X^{\pi}_t],\quad 0 < t < T ,
\end{align}
and, whenever it exists in $\bR\cup\{+\infty\}$, we define
$\bV_T^{\bH}\coloneqq\lim_{t\to T}\bV_t^{\bH}$.

In the following sections we consider some examples of initial enlargements constructed from extending the natural information flow $\bF$ by the addition of the sigma algebra of a random variable $G$ that is measurable in the \hbox{$\sigma$-algebra}~$\cF^R_T$. 
Indeed, we define the  right-continuous enlarged filtration $\bG=\{\cG_t, t\geq0\}$ as follows,
\begin{equation}\label{eq:def.bfg}
    \cG_t \defeq \bigcap_{s>t}\left( \cF_s \vee \sigma(G) \right).
\end{equation}
The additional information provided by the random variable $G$ generates an advantage to a trader that owns it, which we refer to as the $\bG$-agent, with respect to an $\bF$-agent that cannot make use of it.
In order to quantify the value of this advantage, we introduce the additional expected logarithmic utility, as stated in the following definition.
\begin{definition}
The additional expected logarithmic utility of the information of a filtration $\bG \supseteq \bF$ is defined as
\begin{equation*}
 \Delta \bV_T^{\bG} \defeq \bV_T^{\bG} - \bV_T^{\bF}.
\end{equation*}
\end{definition}

We next state a standing assumption, known in the literature as \emph{Jacod's hypothesis}, which helps to ensure that a stochastic decomposition formula is valid in the enlarged filtration $\bG$ for the semimartingales in the filtration~$\bF$.

\begin{assumption}[Jacod's hypothesis]\label{ass.jacod}
 The distribution of the random variable~$G$ is $\sigma$-finite and the regular $\cF_t$-conditional distributions satisfy the following condition
\begin{equation*}
    \PP(G \in \cdot\vert \cF_t) \ll\, \PP(G \in \cdot),\quad \PP-a.s.,\quad 0\leq t < T.
\end{equation*}
\end{assumption}

Given $g\in Supp(G)$, this assumption assures the existence of a jointly measurable $\bF$-adapted process, denoted by 
{$p^g = \prTmenos{p^g}$}, such that 
\begin{equation*}
    p_t^g =  \frac{\PP(G\in dg\vert \cF_t)}{\PP(G\in dg)},\quad 0\leq t<T.
\end{equation*}
For any $g\in Supp(G)$, the process~$p^g$ is an~$\bF$-martingale, 
see~\citet[Lemma~2.1]{AmendingerImkellerSchweizer1998}, and therefore 
for any $\bF$-local martingale $M$, the bracket process 
$\langle M,p^g\rangle^\bF=\prTmenos{\langle M,p^g\rangle^\bF}$
is well-defined.
To simplify notation, we are going to use the following definition
\begin{equation}\label{eq:def.angle.brakets}
\langle M,p^G\rangle_s^\bF \defeq \langle M,p^\cdot\rangle_s^\bF\circ G.
\end{equation}

As previously announced, the reason to ask for Jacod's hypothesis, relies on the following proposition that allows us to compute the $\bG$-semimartingale decomposition of any {$\bF$-local martingale}.
For the proof, we refer to~\citet[Theorem~2.1]{Jacod1985}.
Here, we drop the continuity assumption because the starting filtration is given by the Brownian filtration $\bF$.
\begin{proposition}\label{prop.jacod.semimartingale}
Let $M=\prT{M}$ be an {$\bF$-local martingale} and let $G$ be an {$\cF_T$-measurable} random variable satisfying Jacod's hypothesis. 
Then, the process
\begin{equation}\label{eq:jacod.semimartingale}
    \widehat{M}_t \defeq M_t - \int_0^t \frac{d\langle M,p^G\rangle_s^\bF}{p^G_{s}},\quad 0\leq t < T\ 
\end{equation}
is a $\bG$-local martingale.
\end{proposition}
When the compensator of the process $M$, appearing in~\eqref{eq:jacod.semimartingale}, is absolutely continuous with respect to the Lebesgue measure,
its density is usually called the \emph{information drift}.
We denote the information drift related to the process~$B^R$ as {$\alpha^G=\prTmenos{\alpha^G}$}, which is going to play a crucial role in our computations.
It is defined in the following way
\begin{equation}\label{eq:information.drift}\int_0^t \alpha^G_s ds \defeq \int_0^t \frac{d\langle B^R,p^G\rangle_s^\bF}{p^G_{s}},\quad 0\leq t < T, \end{equation}
and it is such that the process $ W^R_\cdot \defeq B^R_\cdot-\int_0^\cdot\alpha_s^G ds$ is a $\bG$-Brownian motion.

We stress that in~\eqref{eq:jacod.semimartingale}, the interval where the stochastic decomposition is valid, does not include the right extreme $T$ and that, in general, the process $\widehat M$ cannot be extended to the closed interval $[0,T]$.

According to~\citet[Corollaries~2.6 and~2.8]{Amendinger2003}, the semimartingale property is preserved on the whole closed interval if the additional information carried by~$\sigma(G)$, is independent of $\cF_T$ under some probability measure equivalent to~$\PP$ and this is guaranteed by assuming the equivalence version of Jacod's hypothesis in the closed interval $[0,T]$.

\section{Optimization problem}\label{sec:opt.prob}
Combining equations~\eqref{R.SDE} and~\eqref{X.SDE} we get the following system of stochastic differential equations,
	\begin{equation}
	\left \{ \begin{aligned}\label{problemaCentral}
	dX_t^{\pi} &= (1-\pi_t) X_t^{\pi} R_t \, dt + \pi_t X_t^{\pi} \left(\eta_t \, dt + \xi_t \, dB_t^S\right)\\
	dR_t &=  \left[ a_1(t)R_t+a_2(t) \right]dt + b_2(t)dB^R_t\\
	\end{aligned}	\right.
	\end{equation}
that, under the assumption that $\pi\in\cA(\bF)$, gives the evolution of the interest rate process and the portfolio wealth as seen by an $\bF$-agent.

To analyze the same processes when $\pi\in\cA(\bG)$, with $\bG$ defined as in~\eqref{eq:def.bfg},
we employ standard techniques of enlargement of filtrations by looking for the $\bG$-semimartingale decomposition of the pair $(B^R,B^S)$ by employing Proposition~\ref{prop:semimart.repr.bG} below.

By applying It\^o lemma, see also Example~1.5.4.8 in~\cite{chesneyyorjeanblanc2009}, it follows that the process~$R$ admits the following explicit expression, 
\begin{align}\label{eq:solLineal}
 R_t &= \Psi_t  \left[  R_0 + \int_0^t \Psi^{-1}_s a_2(s)  \, ds + \int_0^t \Psi^{-1}_s b_2(s) \, dB^R_s  \right] ,
\end{align}
for $t\in[0,T]$, with  $\Psi_{s,t} \defeq \exp\left({\int_s^t a_1(u) du}\right)$
and $\Psi_t\defeq \Psi_{0,t}$.
The process $R$ is Markovian and Gaussian when $R_0$ is Normal distributed.
The next lemma gives the expression for the conditional distribution of $(R_t \vert  R_s)$, \hbox{$0\leq s \leq t$,} by conveniently handling the explicit expression in~\eqref{eq:solLineal}.
\begin{lemma}\label{lema2}
  Let $R$ be the process defined by~\eqref{R.SDE}.
  For $0\leq s\leq t$, the  conditioned random variable $ (R_t\vert R_s)$ has the following distribution
\begin{equation*}
  (R_t\vert R_s) \disteq
  \cN\left(\Psi_{s,t}R_s  + \Psi_t\int_s^t \Psi^{-1}_u a_2(u) du,\ \Psi_t^2\int_s^t \Psi^{-2}_u b_2^2(u) du \right).
\end{equation*}
\end{lemma}
\begin{proof}The details of the proof are included in  \hyperlink{proof.lem.distr.R}{Appendix \ref*{Appendix}}.
\end{proof}
\subsection{Optimal portfolio}\label{subsec:opt.port.R}
In order to consider a larger class of strategies, that is $\pi\in\cA(\bG)$, we rewrite the SDE~\eqref{problemaCentral} in terms of a $\bG$-Brownian motion and $\bG$-adapted stochastic parameters. 
This is done in the following proposition.
\begin{proposition}\label{prop:semimart.repr.bG}
Under the filtration $\bG$, the processes $X^{\pi}$ and $R$ satisfy the following SDE,
\begin{equation}\label{eq:semimart.repr.bg}
  \left \{ \begin{aligned}
  dX_t^{\pi} &= (1- \pi_t) X_t^{\pi} R_t \, dt +  \pi_t  X_t^{\pi} (\eta_t dt + \xi_t dW_t^S)\\
  dR_t &= \left[a_1(t) R_t + \left(a_2(t) +  b_2(t)\alpha_t^G\right)\right] \, dt + b_2(t) dW_t^R\\
 d W_t^S &\defeq \rho\alpha_t^{G} \, dt+\rho dW_t^R + \sqrt{1-\rho^2}dB_t,\quad 0\leq t<T,
  \end{aligned} \right.
\end{equation}
where $W^R_\cdot=B^R_\cdot-\int_0^\cdot \alpha_s^G$ 
and $B=\prT{B}$ are independent $\bG$-Brownian motions in $[0,T)$.
The additional expected logarithmic utility of the $\bG$-agent is given by the following formula
\begin{equation}\label{eq:add.utility}
    \Delta\bV_T^\bG = \frac{\rho^2}{2}\int_0^T\EE[(\alpha^G_t)^2]dt.
\end{equation}
\end{proposition}

\begin{proof}See \hyperlink{proof.prop.semimart.repr.bG}{Appendix \ref*{Appendix}} for the details of the proof.
\end{proof}

Note that, for a $\bG$-agent, 
the dynamics of $S$ are given by
\begin{equation*}
\frac{dS_t}{S_t} = 
\left(\eta_t + \xi_t\rho\alpha_t^G\right)dt 
+ \xi_t \rho dW_t^R + \xi_t \sqrt{1-\rho^2} dB_t,
\end{equation*}
where $(W^R, B)$ is a bi-dimensional $\bG$-Brownian motion in $[0,T)$.
The expression of the $\bG$-semimartingale $X^\pi$, solution of the SDE~\eqref {eq:semimart.repr.bg}, is given by
\begin{align}\label{X.exponential.solution}
       \frac{X_t^\pi}{x_0} = D_t \exp\Big\{ &\int_0^t  \pi_s(\eta_s-R_s + \rho\xi_s\alpha_s^G - \frac{1}{2}\xi_s^2\pi_s ) ds + \rho\int_0^t \xi_s\pi_s d W_s^R\nonumber\\ 
       &+ \sqrt{1-\rho^2}\int_0^t \xi_s\pi_s dB_s\Big\},\quad 
       0\leq t < T,
\end{align}
where $D$ is defined in~\eqref{D.SDE}.
\begin{remark}
    Note that the market is incomplete because there are infinitely many Equivalent Local Martingale Measures (ELMMs), 
    see \citet[Lemma~4.1 and Remark~4.2]{PhamQuenez2001} for a similar construction of the family of ELMMs.
    This is due to the stochastic nature of the interest rate.
    Nevertheless, when dealing with the logarithmic utility, standard optimization techniques can be applied
    as we show in Example~\ref{ex.log} below. 
    Under a different utility function, 
    one could use the techniques described in~\cite{karatzas1991}.
    \end{remark}
In the following example we solve the optimal portfolio problem under the logarithmic utility and the information flow $\bG$.
\begin{example}\label{ex.log}
    We show here how to solve the optimal portfolio problem in the interval $[0,t]$ with $t<T$, by employing the explicit expression of $\ln X^{\pi}_t$, given in~\eqref{X.exponential.solution}.
    We have that
\begin{equation*}
    \EE \left[\ln \frac{X^{\pi}_t}{x_0} \right] = D_t + \int_0^t \EE\left[ J(\pi_s) \right] ds ,
\end{equation*}
    where $J(\pi) \defeq \pi\left(\eta_s-R_s + \rho\xi_s\alpha_s^G - \frac{1}{2}\xi_s^2\pi \right)$.
    Following the lines of Theorem~16.54 of the monograph~\cite{DiNunnoOksendalProske2009}, for each fixed $(s,\omega)\in[0,t]\times\Omega$, we maximize $J(\pi_s)$, 
    so getting 
    \begin{equation}\label{eq:carteraoptima.log}
        \pi_s^G \defeq \frac{\eta_s-R_s}{\xi_s^2}+\rho\frac{\alpha_s^G}{\xi_s}, \quad 0 \leq s \leq t ,
    \end{equation}
    that is maximum strategy for the optimization problem since $J$ is concave.
    This solution is valid only in the open interval $[0,T)$, and it could be extended  to its closure depending on the integrability conditions of $\alpha^G$, as we show in the following sections.
    \end{example}
    {
    A well-studied case of initial enlargement of filtrations is 
\begin{equation}\label{eq:def.g}
    G = \int_0^T\varphi(t)dB^R_t,
\end{equation}
where $\varphi\in L^2([0,T],dt)$ is deterministic.}
By taking all linear combinations of random variables as in~\eqref{eq:def.g} we have the following set
\begin{equation*}
    \overbar{<B^R_t : 0\leq t\leq T>} = \Big\{  \int_0^T\varphi(t)dB^R_t : \varphi\in L^2([0,T],dt) \Big\} ,
\end{equation*}
where $\overbar{<\cdot>}$ denotes the $L^2(\cF_T^R,\PP)$-closure of the linear span, with $L^2(\cF_T^R,\PP)$
being the set of {$\cF_T^R$-measurable} random variables with finite second moment.
It consists of the set of Gaussian $\cF^R_T$-measurable random variables $G$ such that $\EE[G\vert \cF^R_t]$ is still Gaussian, see Lemma~3.1 in~\cite{Connor2010}.
\begin{remark}
Non-Gaussian random variables, such as $(B^R_T)^2$, do not belong to the above set as their  representation contains a non-deterministic integrand, e.g.
\begin{equation*}
(B^R_T)^2 = T + 2\int_0^T B^R_t dB^R_t.
\end{equation*}
See Subsection~\ref{subsec:non.atomic}, for en explicit example involving the random variable $G=(B_T^R)^2$.
\end{remark}
In the next proposition we give the expression of the information drift defined in Proposition~\ref{prop.jacod.semimartingale} for $G$ as in~\eqref{eq:def.g}.
This problem has already been studied in Theorem~1.1.1 in~\cite{chaleyatJeulin}, see also Example~4.16 in~\cite{Aksamit2017}, 
which deals with the case $T=\infty$.
\begin{proposition}\label{Prop.alpha}
  Let G be as in~\eqref{eq:def.g} with $\int_t^T\varphi^2(s)ds>0$, $\forall t<T$, then the information drift process $\alpha^G=\prTmenos{\alpha^G}$, defined in~\eqref{eq:information.drift} is
  given by
  \begin{equation}\label{eq:alpha_t^g}
    \alpha_t^G= \varphi(t)\frac{\int_t^T\varphi(s)dB^R_s}{\int_t^T\varphi^2(s)ds} ,\quad 0 \leq t < T.
\end{equation}
Moreover, the dynamics of the process $R$ in the enlarged filtration $\bG$ are given by,
\begin{equation}\label{eq:new.dynamics}
   dR_t =  \left[a_1(t) R_t + \left(a_2(t) +  b_2(t)\alpha_t^G\right)\right] dt + b_2(t) dW_t^R,\quad 0\leq t<T. 
\end{equation}
\end{proposition}
Whenever the initial enlargement is generated by the inclusion of the information carried by a not purely atomic random variable, in~\citet[Theorem 4.4]{AmendingerImkellerSchweizer1998}, the additional expected
logarithmic utility of the $\bG$-agent is infinite, that is with $\norm{\varphi}_2\neq 0$,  $\Delta \bV_T^\bG = +\infty$.

\begin{example}
If $\varphi(t) = \Psi_T \Psi^{-1}_t b_2(t)$, the informed agent knows the stochastic part of $R_T$ in~\eqref{eq:solLineal}, so $\sigma(G) = \sigma(R_T)$ and {$\bG=\bF\vee\sigma(R_T)$}.
In this case~\eqref{eq:alpha_t^g} becomes
\begin{equation*}
    \alpha_t^{R_T} = b_2(t) \Psi^{-1}_t \frac{\int_t^T \Psi^{-1}_s b_2(s)dB^R_s}{\int_t^T \Psi^{-2}_s b_2^2(s) ds} 
,\quad 0 \leq t < T.
\end{equation*}
In the more specific case when the interest rate is modeled by an \OU{} process, as defined in~\eqref{Y.SDE}, the  information drift simplifies in
\begin{align}
    \alpha_t^{Y_T} &= \frac{k/\sigma}{\sinh(k(T-t))} \left( ( \mu-Y_t)\ e^{-k(T-t)} - (\mu - Y_T) \right)
    ,\quad 0 \leq t < T.\label{alpha.precise.OU} 
\end{align}
\end{example}
\section{Non-linear type information}\label{sec:non.lin.info}
In this section, we show some examples where the additional information of the {$\bG$-agent} is less precise and not necessarily described by a Gaussian random variable.
Let $G\in L^2(\cF_T^R,\PP)$, then there exists a process $\varphi\in L^2(dt\times\PP,\bF)$ satisfying the Predictable Representation Property (PRP), that is 
\begin{equation}\label{eq:prp.g}
    G = \EE[G] + \int_0^T\varphi_t dB^R_t.
\end{equation}
Note that in~\eqref{eq:prp.g}, the process $\varphi$ may depend on the outcome $\omega\in\Omega$.

Let $B\in \cB(\bR)$ be a Borel set 
and consider the following PRP,
\begin{equation}\label{eq:prp.b}
    \bOne_{\{G\in B\}} = \PP(G\in B) + \int_0^T \varphi_t(B)dB^R_t,
\end{equation}
where, by the predictable process $\varphi(B)=(\varphi_t(B):0\leq t\leq T)$, we denote the unique one within the Hilbert space $L^2(dt\times\PP,\bF)$ that satisfies~\eqref{eq:prp.b} for any fixed $B$ with $\PP(G\in B)>0$.
In Lemma~\ref{lemma.measure.1} below, we prove that $\varphi(\cdot)$ is a vector measure.
\begin{definition}
 A map $\varphi$ from $\cB(\bR)$ to a Banach space $X$ is called a finitely additive vector measure, or simply a vector measure, if whenever $E_1,E_2\in \cB(\bR)$ are disjoint then $\varphi\left(E_1 \cup E_2\right)=\varphi\left(E_1\right)+\varphi\left(E_2\right)$.

If, in addition, $\varphi\left(\bigcup_{n=1}^{\infty} E_n\right)=\sum_{n=1}^{\infty} \varphi\left(E_n\right)$ in the norm topology of $X$ with $E_n\in\cB(\bR),$ 
$\forall n\in\bN$ of pairwise disjoint such that $\bigcup_{n=1}^{\infty} E_n \in \cB(\bR)$, then $\varphi$ is called a countably additive vector measure.
\end{definition}
We refer to~\cite{VectorMeasures77} for more details and a general background on the vector measure theory.
Similarly to  \citet[Definition~4]{VectorMeasures77}, we shall make the following assumption.
\begin{assumption}\label{Assump.bdd.var}
The vector measure $\varphi$ is of bounded variation.
\end{assumption}
\begin{remark}\label{rk.ass.bdd} 
Assumption~\ref{Assump.bdd.var} is trivially satisfied for a discrete random variable as the variation of the vector measure $\varphi$ is bounded above by a constant times the number of values assumed by the random variable.
\end{remark}
The next lemma shows the Radon-Nikodym derivative of the Hilbert valued random measure $\varphi$ with respect to the measure, $\PP^G$, induced by $G$, i.e.~{$\PP^G(\cdot) = \PP(G\in\cdot)$} 
on~$\sigma(G)$.
\begin{lemma}\label{lemma.measure.1}
  The set function $B\longrightarrow \varphi(B)$, with $B\in \cB(\bR)$,
  is a countably additive $L^2(dt\times\PP,\bF)$ valued vector measure and
  there exists a set of processes 
  $\psi = (\psi^g,\, g\in \Spp(G))$ such that
  $\psi^g = \prT{\psi^g}\in L^1(\PP^G)$ satisfying the following relation
\begin{equation}\label{eq:Radon-Nikodym}
    \varphi_t(B) = \int_B \psi_t^g\PP^G(dg),\quad B\in \cB(\bR).
\end{equation}
\end{lemma}
\begin{proof}
    Let $\{B_i\}_{i=1}^\infty\subset \cB(\bR)$ be a sequence of subsets such that 
    \begin{equation*}
        B \defeq \cup_{i=1}^\infty B_i,\quad \emptyset = B_i\cap B_j \ \forall i\neq j\in\bN.
    \end{equation*}
    From \citet[Example~3]{VectorMeasures77} we have that
    {$\bOne_{\{G\in B\}} =  \sum_{i=1}^\infty \bOne_{\{G\in B_i\}}$} holds 
    $\PP$-a.s.\ in {$L^2(dt\times\PP,\bF)$}, therefore,
    \begin{align*}
        \bOne_{\{G\in B\}}  
        =  \sum_{i=1}^\infty \bOne_{\{G\in B_i\}}  
        = \PP^G(B) + \sum_{i=1}^\infty \int_0^T \varphi_t(B_i)dB^R_t 
        =  \int_0^T \sum_{i=1}^\infty\varphi_t(B_i)dB^R_t.
    \end{align*}
    In the second equality, we applied the additive property of the probability measure~$\PP$ while, in the third one,
    since $\varphi(B)\in L^2(dt\times\PP,\bF)$ for any $B\in \cB(\bR)$, 
    we apply the stochastic Fubini theorem.
    From the uniqueness of the representation, we deduce that $\varphi(B) = \sum_{i=1}^\infty\varphi(B_i)$,
    and we conclude that $\varphi$ is an $L^2(dt\times\PP,\bF)$-vector measure.
    As $\varphi \ll \PP^G$ on $\sigma(G)$, the last claim of the lemma follows
    by applying Proposition 2.1 in~\cite{Kakihara11}.
\end{proof}
Let $g\in Supp(G)$, by Lemma~\ref{lemma.measure.1} we know that 
there exists a process $\psi^g$ such that,
\begin{equation}\label{eq:repr.binary}
    \bOne_{\{G\in dg\}} =
    \PP^G(dg) + \int_0^T \psi_s^g\PP^G(dg) dB^R_s. 
\end{equation}
When $G$ is purely atomic,~\eqref{eq:repr.binary} reduces to
$\bOne_{\{G = g\}} =
    \PP^G(g) + \int_0^T \psi_s^g\PP^G(g) dB^R_s.$
By employing the Radon-Nikodym derivative introduced in  Lemma~\ref{lemma.measure.1}, we can get a more explicit expression of the information drift as the next lemma shows.

The following result is already known, see for example Theorem 2.6 in~\cite{AnkirchnerImkeller06}. 
\begin{lemma}\label{lemma.drift.general}
  Let $G\in L^2(\cF_T^R,\PP)$, then the information drift $\alpha^G$ has the following expression
  \begin{equation}\label{eq:inf.drift.vect.meas}
        \alpha_t^G =  \frac{\psi^G_t}{p_{t-}^G},\quad 0 \leq t < T .
    \end{equation}
\end{lemma}
The information drift in~\eqref{eq:inf.drift.vect.meas} gets an explicit form as soon as one can compute the Radon-Nikodym derivative appearing in~\eqref{eq:Radon-Nikodym}. 
For example, we next give the expression of the information drift when $G$ is a Bernoulli random variable. \begin{theorem}\label{alpha.binary}
  Let $\Spp(G)=\{0,1\}$, then
  \begin{equation}\label{eq:alpha.g.{0,1}}
      \alpha_t^G =(-1)^{G+1} \varphi_t\frac{G-\EE[G\vert \cF_t]}{\VV[G\vert \cF_t]},\quad 0\leq t < T.
  \end{equation} 
\end{theorem}
\begin{proof}
Rewriting equation~\eqref{eq:repr.binary} for the discrete case as
$\bOne_{\{G=g\}} = \PP^G(g) + \int_0^T\varphi^g_t dB_t^R$, $g\in\{0,1\}$,
it follows that $\varphi_t = \varphi^1_t$ as $G = \bOne_{\{G=1\}}$.
As $\bOne_{\{G=0\}}+\bOne_{\{G=1\}} = 1$ 
we conclude that $\varphi^1_t = -\varphi^0_t$.
The result follows since
$\EE[G\vert \cF_t] = \PP(G=1\vert \cF_t)$ and~$\VV[G\vert \cF_t] = \PP(G=1\vert \cF_t)(1-\PP(G=1\vert \cF_t))$.
\end{proof}
\subsection{Half-bounded interval}\label{subsec:infinite.interval}
In this section we assume that the $\bG$-agent knows 
if $R_T$ is going to be less or greater than a given value $c\in\bR$.
To this aim, we consider the random variable $A \defeq \bOne \{ R_T\leq c \}$, and the enlarged filtration 
${\bG} = \bF\vee \sigma(A)\ $.

Similar kind of information, but referring to the risky asset, was proposed by~\citet[Example 4.6]{karatzas1996}.
\begin{proposition}\label{drift.interval.infinite}
    Let $A = \bOne \{ R_T\leq c \}$, then
    \begin{equation}\label{eq:alpha.interval.infinite}
       \alpha^a_{t} 
    = (-1)^{a} \Psi_T \Psi^{-1}_t b_2(t) \frac{ f_{R_T\vert R_t}(c\vert R_t)}{\PP( A=a \, \vert \cF_t)}
    , \quad a\in\{0,1\},\quad t < T, 
    \end{equation}
    where, by $f_{R_T\vert R_t}(c\vert R_t)$, we denote the Gaussian density whose parameters are given in Lemma~\ref{lema2}.
\end{proposition}
\begin{proof}
    The proof follows a standard argument, however it is adapted to the explicit expression, \eqref{eq:solLineal}, available for affine processes. See \hyperlink{proof.drift.interval.infinite}{Appendix~\ref*{Appendix}} for the details.
\end{proof}
Expression~\eqref{eq:alpha.interval.infinite} simplifies for the Vasicek model to the following one
    \begin{equation*}
       \alpha^a_{t} 
    = (-1)^{a}\sigma e^{-k(T-t)}\frac{ f_{Y_T\vert Y_t}(c\vert Y_t)}{\PP( A=a \, \vert \cF_t)}
    , \quad a\in\{0,1\},\quad t < T. 
    \end{equation*}

\begin{corollary}
The random variable $\bOne\{ R_T\leq c \}$ admits the following predictable representation,
  \begin{equation*}
      \bOne \{ R_T\leq c \} = \PP( R_T\leq c) - \Psi_T\int_0^T \Psi^{-1}_t b_2(t) f_{R_T\vert R_t}(c\vert R_t)dB^R_t.
  \end{equation*}
  In particular, when {$A=\bOne \{ B^R_T\leq c \}$}, 
we get the following expression,
\begin{equation*}
    \bOne \{ B^R_T\leq c \} = \PP( B^R_T\leq c) - \int_0^T \frac{1}{\sqrt{2\pi(T-t)}}\exp\left(-\frac{(c-B^R_t)^2}{2(T-t)}\right) dB^R_t.
\end{equation*}
\end{corollary}
\begin{proof}
The result follows by using~\eqref{eq:repr.binary} and  applying Theorem~\ref{alpha.binary}.
\end{proof}

From Proposition~\ref{prop.jacod.semimartingale}, there exists a ${\bG}$-Brownian motion $W^R$ such that
\begin{equation*}
    dR_t = \left[a_1(t) R_t + \left(a_2(t) +  b_2(t)\alpha_t^A\right)\right] \, dt + b_2(t) dW_t^R,\quad t<T.
\end{equation*}

The above expression, together with the arguments of Subsection~\ref{subsec:opt.port.R},
allows us to write the dynamics of the portfolio 
under the information flow ${\bG}$ in a way similar to the one shown in Proposition~\ref{prop:semimart.repr.bG}.
However, in this case the additional expected
logarithmic utility of the $\bG$-agent is finite, in agreement with the results in \citet[Theorem 4.1]{AmendingerImkellerSchweizer1998}.

\subsection{Bounded interval}\label{subsec:finite.interval}
In this section, we repeat the same computations of the previous sections assuming that the trader knows if $R_T$ is going to belong or not to a certain bounded interval, that is ${\bG} = \bF\vee \sigma(L)$ with 
$L \defeq \bOne \{ c_1\leq R_T\leq c_2 \}$.

We have that the information drift is equal to
\begin{equation*}
       \alpha^l_{t} 
    = (-1)^l \Psi_T \Psi^{-1}_t b_2(t) \frac{f_{R_T\vert R_t}(c_2\vert R_t)-f_{R_T\vert R_t}(c_1\vert R_t) }{\PP( L=l \vert \cF_t)}
    ,\quad t < T. 
\end{equation*}
or, for \hbox{$L \defeq \bOne \{ c_1\leq Y_T\leq c_2 \}$}, to
\begin{equation}\label{eq:alpha.interval.finite.ou}
       \alpha^l_{t} 
    = (-1)^{l}\sigma e^{-k(T-t)}\frac{ f_{Y_T\vert Y_t}(c_2\vert Y_t)-f_{Y_T\vert Y_t}(c_1\vert Y_t)}{\PP( L=l \, \vert \cF_t)}
    , \quad t < T. 
    \end{equation}

The random variable $\bOne\{ c_1\leq R_T\leq c_2 \}$ admits the following representation,
\begin{align*}
     \bOne \{ c_1\leq R_T\leq c_2 \} =& P(c_1\leq R_T\leq c_2 ) - \Psi_T\int_0^T \Psi^{-1}_t b_2(t) f_{R_T\vert R_t}(c_2\vert R_t) dB_t^R\\
     &+\Psi_T\int_0^T \Psi^{-1}_t b_2(t) f_{R_T\vert R_t}(c_1\vert R_t) dB_t^R.
\end{align*}

Also in this case, ${\Delta \bV_T^\bG}<\infty$.

\subsection{Non-atomic information}\label{subsec:non.atomic}
In this section, we consider the enlargement of filtration ${\bG} = \bF\vee \sigma(G)$
where $G = \left(B_T^R\right)^n$.
\begin{proposition}\label{prop.drift.square.BM}
    Let $G = \left(B_T^R\right)^n$, then for $0\leq t < T$,
        \begin{equation}\label{eq:drift.square.bm}
        \alpha^G_t = \left\{ \begin{array}{lcc}
            \dfrac{\sqrt[n]{G}\tanh\left(\dfrac{\sqrt[n]{G}B_t^R}{T-t}\right)-B_t^R}{T-t} &  \text{ if $n$ is even} & \\
             \\ \dfrac{\sqrt[n]{G}-B_t^R}{T-t} &   \text{ if $n$ is odd}  & \\
             \end{array}
   \right.
    \end{equation}
\end{proposition}
\begin{proof}
When $n$ is odd we recover the Brownian bridge and the result is well-known.
When it is even we solve the case $n=2$ and
the rest can be deduced by raising $G$ to $2/n$.
We proceed by applying Lemma~\ref{lemma.drift.general}.
To this end, we apply well-known results on the Brownian bridge.
\begin{align}\label{eq.repr.square.BM}
\bOne_{\{(B^R_T)^2\in dg\}} &= 
\bOne_{\{B^R_T\in \sqrt{dg}\}} + \bOne_{\{B^R_T\in -\sqrt{dg}\}} \notag\\
&=  \PP((B^R_T)^2\in dg) + 
\int_0^T \left( \psi^{\sqrt{g}}_s\PP(B^R_T\in dg)+\psi^{-{\sqrt{g}}}_s\PP(B^R_T\in -dg) \right) dB^R_s,
\end{align}
where $\psi$ is the Radon-Nikodym derivative appearing in Lemma~\ref{lemma.measure.1} with respect to the random variable $B_T^R$,
for which the information drift is known.
Then,
\begin{align*}
     \psi^{m}_t\PP(B^R_T\in dm) &= \PP(B_T^R \in dm \vert \cF_t)\frac{m-B^R_t}{T-t}, \quad m\in\bR,
\end{align*}
and finally, by using~\eqref{eq.repr.square.BM} and Lemma~\ref{lemma.drift.general} with respect to $G=(B_T^R)^2$ we get,
\begin{align*}
    \alpha^g_t &= \frac{
    \PP(B^R_T\in \sqrt{dg} \vert \cF_t)\frac{{\sqrt{g}}-B^R_t}{T-t} + \PP(B^R_T\in -\sqrt{dg} \vert \cF_t) \frac{-{\sqrt{g}}-B^R_t}{T-t} }{\PP((B_T^R)^2\in dg\vert \cF_t)}\\
    &= \frac{\exp\left(-\frac{(\sqrt{g}-B_t^R)^2}{2(T-t)}\right)\frac{{\sqrt{g}}-B^R_t}{T-t} + \exp\left(-\frac{(-\sqrt{g}-B_t^R)^2}{2(T-t)}\right) \frac{-{\sqrt{g}}-B^R_t}{T-t} }{\exp\left(-\frac{(\sqrt{g}-B_t^R)^2}{2(T-t)}\right)+\exp\left(-\frac{(-\sqrt{g}-B_t^R)^2}{2(T-t)}\right)}\\
    &
    =  \frac{\sqrt{g}\tanh\left(\dfrac{\sqrt{g}B_t^R}{T-t}\right)-B_t^R}{T-t},
\end{align*}
where $g>0$ 
and the result follows true.
\end{proof}
From Theorem 4.4 of~\cite{AmendingerImkellerSchweizer1998}, ${\Delta \bV_T^\bG} = +\infty$ whenever $\rho\neq 0$.

\section{Markovian modulation}\label{sec:markov.modul}
In this section we introduce a discrete-time Markov chain $\varepsilon$,
which modulates the dynamics of the interest rate.
Let $E=\{0,1\}$ be its state space and denote by~$\bF^\varepsilon$ its natural filtration.
To embed the discrete time process $\varepsilon$ in a continuous time model, 
we assume that $\varepsilon$ is allowed to change its value only on the fixed time set $\{t_0 <t_1 <\ldots<t_n=T\}$.

By introducing the stochastic modulation in our model, we are going to assume that the dynamics of the assets are given by the following SDEs,
\begin{align}
    dD_t &= {D_t}\,Y^{\varepsilon}_t\, dt, \quad D_0 = 1 \label{SDE.e.D}\\
    dS_t &= {S_t}\left(\eta_t\, dt + \xi_t\, dB^S_t\right),\quad S_0 = s_0 >0, \label{SDE.e.S}
\end{align}
so that the short rate can switch between two different processes, namely~$\{Y^0,Y^1\}$, where 
\begin{align}\label{mod.OU}
    dY^e_t = k^e(\mu^e-Y^e_t)dt + \sigma dB_t^Y,\quad t_\kappa\leq t < t_{\kappa+1} ,
\end{align}
satisfying the continuity condition $Y^e_{t_\kappa}\defeq \lim_{t\to t_\kappa^-}Y^{\varepsilon_{t_{\kappa-1}}}_t$.

To make the notation consistent with the one used for the continuous-time processes, 
we set, in~\eqref{SDE.e.D} and~\eqref{SDE.e.S}, 
$\varepsilon_t \defeq \varepsilon_{t_\kappa}$ for $t_\kappa\leq t < t_{\kappa+1}$.

If we assume that $\bF^\varepsilon\subset\bF$, 
then the classical It\^o integral with respect to $B^S$ is well-defined, and we recover a modulated version of the Merton problem, discussed already in~\cite{Bauerle}.
If we consider $\varepsilon$ independent of $(B^S,B^Y)$, then the classical It\^o integral continues to be well-defined, see~\citet[Section 3.3]{OksendalSDE}. 
However, the new source of randomness generates incompleteness in the market. 
This problem can be seen as a special case of the one dealt with in~\cite{karatzas1991} that has been partially solved, for instance, in~\cite{GhoshGoswamiKumar2009} and in~\cite{SotomayorCadenillas2009}.

Here we are going to generalize the above results, by assuming that {$\varepsilon_t\in\cF^Y_{t_{\kappa+1}}\setminus\cF^Y_{t_\kappa}$}.

The novelty, then, consists of including some anticipating information in the market as a consequence of the knowledge of $\varepsilon$.
In this case, if the strategy $\pi$ is $\bF\vee\bF^\varepsilon$-adapted, we may have that the It\^o integral $\int_0^T\pi_t\xi_t dB_t^S$ is not well-defined.
The enlargement $\bG\supset\bF$, 
which incorporates the information of $\varepsilon$ into $\bF$, is a sequence of initial enlargements 
given by
\begin{equation}\label{eq:g.markov}
    \cG_t = \bigcap_{s>t}\left( \cF_s \vee \sigma(\varepsilon_{t_0}, \varepsilon_{t_1}, \ldots, \varepsilon_{t_\kappa}) \right),\quad t_\kappa \leq t < t_{\kappa+1},\quad \kappa<n.
\end{equation}

We can reduce this problem to the framework of initial enlargements by considering a sequence of problems, one for each interval $[t_\kappa,t_{\kappa+1})$.
We still impose that Assumption~\ref{ass.jacod} holds true on $\varepsilon_{t_\kappa}$, for any $\kappa < n$, 
so that there exists a family of jointly measurable processes~$p^{e,\kappa}$ such that
\begin{equation*}
    p_t^{e,\kappa} =  \frac{\PP( \varepsilon_{t_\kappa} = e \vert \cF_t)}{\PP(\varepsilon_{t_\kappa} = e)},
    \quad e\in\{0,1\},\quad \kappa\in\{0,1,\dots,n-1\},\quad t_\kappa\leq t< t_{\kappa+1}. 
\end{equation*}
Having assumed that $\varepsilon_{t_\kappa}\in\cF^Y_{t_{\kappa+1}}$, the following PRP holds
\begin{equation}\label{eq:prp.g.varep}
    \varepsilon_{t_\kappa} = \PP(\varepsilon_{t_\kappa} = 1) + \int_0^{t_{\kappa+1}}\varphi^\kappa_t dB^Y_t.
\end{equation}
\begin{proposition}\label{Markov.infor.drift}
With the enlargement of filtrations defined in~\eqref{eq:g.markov}, we have that the information drift is given by
$\alpha^\varepsilon_t \defeq \alpha^\varepsilon_t(\kappa)$, as $t_\kappa\leq t < t_{\kappa+1}$, where
\begin{equation}\label{eq:alpha.varepsilon}
    \alpha_t^\varepsilon(\kappa) = {(-1)^{\varepsilon_{t_\kappa}+1}}
\varphi_t^\kappa \frac{\varepsilon_{t_\kappa}-\EE[\varepsilon_{t_\kappa}\vert \cF_t]}
{\VV[\varepsilon_{t_\kappa}\vert \cF_t]}.
\end{equation}
and the processes $X^{\pi}$ and $Y$ satisfy the following SDE, as  $t_\kappa\leq t<t_{\kappa+1}$,
\begin{equation}\label{eq:varepsilon.SDE}
  \left \{ \begin{aligned}
  dX_t^{\pi} &= (1- \pi_t) X_t^{\pi} Y^\varepsilon_t \, dt +  \pi_t  X_t^{\pi} (\eta_t dt + \xi_t dW_t^S)\\
  dY^\varepsilon_t &= \left[k^\varepsilon(\mu^\varepsilon-Y^\varepsilon_t) +  \sigma \alpha_t^\varepsilon(\kappa)\right] \, dt + \sigma dW_t^Y\\
 d W_t^S &\defeq \rho\alpha_t^\varepsilon(\kappa) \, dt+\rho dW_t^R + \sqrt{1-\rho^2}dB_t,
  \end{aligned} \right.
\end{equation}
where $W^Y$ and $B$ are independent $\bG$-Brownian motions on $[t_\kappa,t_{\kappa+1})$.
\end{proposition}
\begin{proof}
Let $t_\kappa\leq t < t_{\kappa+1}$,
by applying Jacod's hypothesis and Theorem~\ref{alpha.binary}
we know that $B_\cdot^Y-\int_{t_\kappa}^\cdot \alpha_s^\varepsilon(\kappa)ds$ is a $\bG$-Brownian motion on $[t_\kappa,t_{\kappa+1})$.
Moreover, as we are considering only enlargements with purely atomic random variables, 
by Theorem~4.1 in~\cite{AmendingerImkellerSchweizer1998}, 
we have that
\begin{equation*}
    \frac{1}{2}\int_0^T\EE\left[\left(\alpha^\varepsilon_t\right)^2 \right]dt = -\sum_{\kappa=0}^{n-1} H(\varepsilon_{t_\kappa}) <+\infty,
\end{equation*}
being $H(\varepsilon_{t_\kappa})$ the entropy operator of the random variable $\varepsilon_{t_\kappa}$.
Then, we can apply Theorem 3.2 in~\cite{Imkeller2003} to get the result up to $T$ included.
\end{proof}
In the following, we provide the additional expected logarithmic gains when the anticipating information is related to the modulated stochastic interest rate.
\begin{theorem}
Under the previous set-up,
\begin{align}\label{eq.gains.B}
    \Delta\bV_T^\bG=& \frac{1}{2}\int_0^T\EE\left[\left(\alpha_t^\varepsilon\right)^2\right] dt - \sum_{\kappa=0}^{n-1}\int_{t_\kappa}^{t_{\kappa+1}}
\EE\left[\frac{\varphi_t^\kappa}{\xi_t}
 \frac{\int_t^{t_{\kappa+1}}\varphi_s^\kappa\phi_s^\kappa ds}
{\VV[\varepsilon_{t_\kappa}\vert \cF_t]} \right] dt,
\end{align}
for some $\bF^Y$-adapted process $\phi^\kappa=(\phi^\kappa_s,\, t_\kappa\leq s \leq t_{\kappa+1})$.
\end{theorem}
\begin{proof}
By standard arguments, for example see \citet[Theorem 2.13]{AnkirchnerImkeller05}, the gains under the filtration $\bG$ at $T$ are given by
\begin{equation*}
    \bV_T^\bG
    = \bV_T^\bF  + \frac{1}{2}\int_0^T \EE\left[\left(\alpha_t^\varepsilon \right)^2\right]dt  +\int_0^T \EE\left[\frac{\eta_{\varepsilon_t} - Y_t^\varepsilon}{\xi_{\varepsilon_t}}\alpha_t^\varepsilon \right]dt.
\end{equation*}
However, due to the Markovian modulation, a new term appears in the expression and we can rewrite it in the following way,
\begin{align*}
    \int_0^T\EE\left[\frac{\eta_t - Y_t^\varepsilon}{\xi_t}\alpha_t^\varepsilon \right] dt 
    =& \sum_{\kappa=0}^{n-1}\int_{t_\kappa}^{t_{\kappa+1}}\EE\left[\frac{\eta_t - Y_t^\varepsilon}{\xi_t}\alpha_t^\varepsilon(\kappa) \right] dt\\
    =& \sum_{\kappa=0}^{n-1}\int_{t_\kappa}^{t_{\kappa+1}}\EE\left[\frac{\eta_t - Y_t^\varepsilon}{\xi_t}{(-1)^{\varepsilon_{t_\kappa}+1}}
    \varphi_t^\kappa \frac{\varepsilon_{t_\kappa}-\EE[\varepsilon_{t_\kappa}\vert \cF_t]}
    {\VV[\varepsilon_{t_\kappa}\vert \cF_t]} \right] dt\\
    =& \sum_{\kappa=0}^{n-1}\int_{t_\kappa}^{t_{\kappa+1}}\EE\left[\frac{\eta_t - Y_t^\varepsilon}{\xi_t}{(-1)^{\varepsilon_{t_\kappa}+1}}
    \varphi_t^\kappa \frac{\int_t^{t_{\kappa+1}}\varphi_s^\kappa dB_s^Y}
    {\VV[\varepsilon_{t_\kappa}\vert \cF_t]} \right] dt\\
    =& -\sum_{\kappa=0}^{n-1}\int_{t_\kappa}^{t_{\kappa+1}}\EE\left[\frac{Y_t^\varepsilon}{\xi_t}{(-1)^{\varepsilon_{t_\kappa}+1}}
    \varphi_t^\kappa \frac{\int_t^{t_{\kappa+1}}\varphi_s^\kappa dB_s^Y}
    {\VV[\varepsilon_{t_\kappa}\vert \cF_t]} \right] dt .
    \end{align*}
By using the following PRP,
    \begin{equation*}
        Y_t^\varepsilon(-1)^{\varepsilon_{t_\kappa}+1} =
        \EE[Y_t^\varepsilon(-1)^{\varepsilon_{t_\kappa}+1}] + \int_0^{t_{\kappa+1}} \phi_s^\kappa dB_s^Y ,
    \end{equation*}
we obtain
    \begin{align*}
    \int_0^T\EE\left[\frac{\eta_t - Y_t^\varepsilon}{\xi_t}\alpha_t^\varepsilon \right] dt 
    =& -\sum_{\kappa=0}^{n-1}\int_{t_\kappa}^{t_{\kappa+1}}\EE\left[\frac{Y_t^\varepsilon}{\xi_t}{(-1)^{\varepsilon_{t_\kappa}+1}}
    \varphi_t^\kappa \frac{\int_t^{t_{\kappa+1}}\varphi_s^\kappa dB_s^Y}
    {\VV[\varepsilon_{t_\kappa}\vert \cF_t]} \right] dt\\
=& - \sum_{\kappa=0}^{n-1}\int_{t_\kappa}^{t_{\kappa+1}}
\EE\left[\frac{\varphi_t^\kappa}{\xi_t}
 \frac{\int_t^{t_{\kappa+1}}\varphi_s^\kappa\phi_s^\kappa ds}
{\VV[\varepsilon_{t_\kappa}\vert \cF_t]} \right] dt.
\end{align*}
\end{proof}

The expression~\eqref{eq.gains.B} is hard to compute in general because it follows by an existence result given by the PRP property.
In the next corollary, we show the result for a simpler but more explicit case.
\begin{corollary}
Under the previous set-up, if $\{Y_0,Y_1\}$ are constant values, then
\begin{align}
    \Delta\bV_T^\bG=& \frac{1}{2}\int_0^T\EE\left[\left(\alpha_t^\varepsilon\right)^2\right] dt + 
(Y_1-Y_0)\sum_{\kappa=0}^{n-1}\int_{t_\kappa}^{t_{\kappa+1}}\EE\left[\frac{\varphi_t^\kappa}{\xi_t}
\right] dt.
\end{align}
\end{corollary}
\begin{proof}
Under the assumption that $\{Y_0,Y_1\}$ are constant values, 
we notice that
\begin{equation*}
    \frac{\eta_t-Y^\varepsilon_t}{\xi_t} =\frac{\eta_t}{\xi_t} - \frac{Y_0}{\xi_t} + \varepsilon_{t_\kappa}\frac{Y_1-Y_0}{\xi_t},\quad t_\kappa\leq t < t_{\kappa+1}.
\end{equation*}
Then we get,
\begin{align*}
    \int_0^T\EE\left[\frac{\eta_t - Y^\varepsilon_t}{\xi_t}\alpha_t^\varepsilon \right] dt 
    =& \sum_{\kappa=0}^{n-1}\int_{t_\kappa}^{t_{\kappa+1}}\EE\left[\varepsilon_{t_\kappa}\frac{Y_1-Y_0}{\xi_t}\alpha_t^\varepsilon(\kappa) \right] dt\\
=&(Y_1-Y_0)\sum_{\kappa=0}^{n-1}\int_{t_\kappa}^{t_{\kappa+1}}\EE\left[\frac{\varepsilon_{t_\kappa}}{\xi_t}
\varphi_t^\kappa \frac{\int_t^{t_{\kappa+1}}\varphi_s^\kappa dB_s^Y}
{\VV[\varepsilon_{t_\kappa}\vert \cF_t]} \right] dt\\ 
=& 
(Y_1-Y_0)\sum_{\kappa=0}^{n-1}\int_{t_\kappa}^{t_{\kappa+1}}\EE\left[\frac{\varphi_t^\kappa}{\xi_t}
 \frac{\left(\int_t^{t_{\kappa+1}}\varphi_s^\kappa dB_s^Y\right)^2}
{\VV[\varepsilon_{t_\kappa}\vert \cF_t]} \right] dt \\
=&
(Y_1-Y_0)\sum_{\kappa=0}^{n-1}\int_{t_\kappa}^{t_{\kappa+1}}\EE\left[\frac{\varphi_t^\kappa}{\xi_t}
\right] dt.
\end{align*}
In the last equality we used that
\begin{align*}
    \VV[\varepsilon_{t_\kappa}\vert \cF_t] =& \EE[(\varepsilon_{t_\kappa}-\EE[\varepsilon_{t_\kappa}\vert \cF_t])^2\vert\cF_t]  =  
    \EE\left[\left( 
    \int_t^{t_{\kappa+1}} \varphi_s^\kappa dB^Y_s
    \right)^2\vert\cF_t\right].
\end{align*}
The result then follows by known properties of the It\^o integral.
\end{proof}
\begin{example}\label{Ex.min}
Let $m=\prT{m}$ be the running minimum of $B^Y$, i.e.,
\begin{equation*}
    m_t\defeq\inf_{0\leq s\leq t} B^Y_s.
\end{equation*}
We use the Markov property of $B^Y_\cdot-m_\cdot$ in order to define $\varepsilon$ as follows,
\begin{equation}\label{eq:ex.drawdown}
    \varepsilon_{t_\kappa} = \bOne_{ \{B^Y_{t_{\kappa+1}}-m_{t_{\kappa+1}}\leq  c \}}  ,\quad c\in\bR^+,\quad \kappa\in\{0,1,\dots,n-1\}.
\end{equation}
We compute the information drift $\alpha^\varepsilon$ when 
$\{\varepsilon_{t_\kappa} = 1\}$ by computing the following process,
\begin{equation}\label{eq:pt1k}
    p_t^{1,\kappa} = \frac{\PP( B^Y_{t_{\kappa+1}} - m_{t_{\kappa+1}}\leq c \vert\cF_t)}{\PP(B^Y_{t_{\kappa+1}} - m_{t_{\kappa+1}} \leq c)}.
\end{equation}
The process $\alpha^\varepsilon(\kappa)$ is strongly related with the $\bF$-local martingale $p^{e,k}$.
Indeed, in \citet[Remark 5]{baudoin2003}, we can find the following relationship
\begin{equation}\label{eq:Baudoin.represent}
    dp_t^{e,\kappa} = p_t^{e,\kappa} \alpha_t^e(\kappa) dB^Y_t.
\end{equation}
We operate with the numerator, looking for a more explicit expression.
We denote by
$m_{s,t}\defeq\inf_{u\in[s,t]} B^Y_u$ and $a\wedge b\defeq\min\{a,b\}$, then
\begin{align}\label{Ex.min.prob1}
    \PP(  B^Y_{t_{\kappa+1}} - m_{t_{\kappa+1}} \leq c \vert\cF_t) 
    &= \PP( B^Y_{t_{\kappa+1}} - m_{t_{\kappa+1}} \leq c,\, m_t\leq m_{t,t_{\kappa+1}}\vert\cF_t) \notag\\
    & \quad +\PP( B^Y_{t_{\kappa+1}} - m_{t_{\kappa+1}} \leq c,\, m_t > m_{t,t_{\kappa+1}}\vert\cF_t) \notag\\
    &= \PP(Y_{t_{\kappa+1}-t} \geq K_t-c ,\, X_{t_{\kappa+1}-t}\geq {K_t}) \notag\\
    & \quad + \PP(X_{t_{\kappa+1}-t}-Y_{t_{\kappa+1}-t} \leq c ,\, X_{t_{\kappa+1}-t}\leq {K_t})
\end{align}
where $K_t \defeq B^Y_t-m_t$, $X_\tau \defeq B^Y_{t_{\kappa+1}-\tau}-m_{t_{\kappa+1}-\tau,t_{\kappa+1}}$ and $Y_\tau \defeq  B^Y_{t_{\kappa+1}-\tau} - B^Y_{t_{\kappa+1}}$.
From Proposition~3.2.1.1 of~\cite{chesneyyorjeanblanc2009}, the vector $(X_\tau,Y_\tau)$ has the following density function,
\begin{align*}
p(x,y,\tau) 
&= \frac{2(y-2x)}{\sqrt{2\pi \tau^3}} \exp\left(-\frac{(2x-y)^2}{2\tau}\right), 
\end{align*}
so we can rewrite~\eqref{Ex.min.prob1} as,
\begin{align*}
    \PP(B^Y_{t_{\kappa+1}}-m_{t_{\kappa+1}}\leq c \vert\cF_t) &= 
    \int_{K_t}^{+\infty}\int^{+\infty}_{{K_t}-c} p(x,y,t_{\kappa+1}-t) dydx \\
    &\quad + \int^{K_t}_{0}\int^{+\infty}_{x-c} p(x,y,t_{\kappa+1}-t) dydx\\
    &=2\int_{K_t}^{+\infty}  
    \frac{\Phi'\left( \frac{2x+c-K_t}{\sqrt{t_{\kappa+1}-t}} \right)}{\sqrt{t_{\kappa+1}-t}} dx+ 2\int^{K_t}_{0}\frac{\Phi'\left( \frac{x+c}{\sqrt{t_{\kappa+1}-t}} \right)}{\sqrt{t_{\kappa+1}-t}} dx \\
    &={\Phi}\left(\frac{c-K_t}{\sqrt{t_{\kappa+1}-t}}\right) - 
    \overline{\Phi}\left(\frac{c+K_t}{\sqrt{t_{\kappa+1}-t}}\right).
\end{align*}
By letting $f(t,K_t)\defeq \PP(B^Y_{t_{\kappa+1}} - m_{t_{\kappa+1}}  \leq c \vert\cF_t)$, we aim to compute $df(t,K_t)$ via It\^o's lemma.
Since the partial derivatives satisfy the following equation,
    \begin{align*}
        \frac{\partial}{\partial K}f(t,K_t) &= \frac{1}{\sqrt{t_{\kappa+1}-t}}\Phi'\left(\frac{c-K_t}{\sqrt{t_{\kappa+1}-t}}\right)-\frac{1}{\sqrt{t_{\kappa+1}-t}}\Phi'\left(\frac{c+K_t}{\sqrt{t_{\kappa+1}-t}}\right)\\
       0 &= \frac{\partial}{\partial t}f(t,K_t)+\frac{1}{2} \frac{\partial^2}{\partial K^2}f(t,K_t) ,
    \end{align*}
    we conclude that
    $df(t,K_t) = \frac{\partial}{\partial K}f(t,K_t) dK_t.$
As the process $M$ is increasing only in the set $\{t,\,K_t = 0\}$, 
it follows that $\frac{\partial}{\partial K}f(t,0)dM_t = 0$,
and the following relationship holds,
\begin{equation*}
    df(t,K_t)
    = - \frac{\partial}{\partial K}f(t,K_t) dB^Y_t = \frac{\Phi'\left(\frac{c+K_t}{\sqrt{t_{\kappa+1}-t}}\right) - \Phi'\left(\frac{c-K_t}{\sqrt{t_{\kappa+1}-t}}\right)}{\sqrt{t_{\kappa+1}-t}} dB^Y_t.
\end{equation*}
Going back to $\eqref{eq:pt1k}$, we conclude that the following representation holds,
\begin{equation*}
    dp_t^{1,\kappa} = p_t^{1,\kappa} \frac{\Phi'\left(\frac{c+K_t}{\sqrt{t_{\kappa+1}-t}}\right) - \Phi'\left(\frac{c-K_t}{\sqrt{t_{\kappa+1}-t}}\right)}{\sqrt{t_{\kappa+1}-t}\,\PP(\varepsilon_{t_\kappa} =  1 \vert\cF_t)} d B^Y_t ,
\end{equation*}
        and, by using~\eqref{eq:Baudoin.represent}, we get the result for the case $e=1$.
        Repeating the same for $e=0$, we have that, for $t_\kappa\leq t < t_{\kappa+1}$,
        \begin{equation}\label{eq:drift.markov}
            \alpha^e_t(k) = 
       \frac{(-1)^e}{\sqrt{t_{\kappa+1}-t}}
       \frac{\Phi'\left(\frac{c+B^Y_t-m_t}{\sqrt{t_{\kappa+1}-t}}\right) - 
        \Phi'\left(\frac{c+m_t- B^Y_t}{\sqrt{t_{\kappa+1}-t}}\right)}{\PP(\varepsilon_{t_\kappa} = e\vert\cF^Y_t)},\quad e\in\{0,1\}.
        \end{equation}
\end{example}

\section{Numerical example}\label{sec:num.illustr}
In this section, we present a numerical experiment that illustrates the results obtained in the previous sections. We set the time horizon to $T=1$ and discretize the time interval into $n=100$ steps, referred to as trading periods. For the dynamics of the assets, we specify the market coefficients as follows:
\begin{equation*}
dD_t = D_t \, Y_t \, dt, \quad 
dS_t = S_t \, \left(0.05\ dt + 0.01\ dB^S_t\right),
\end{equation*}
where we assumed the values $\eta_t = 0.05$ and $\xi_t = 0.01$, for all $0\leq t\leq 1$.
The process $B^S=\prs{B^S}$ is a standard Brownian motion. 
With respect to the interest rate, we assume that $Y = \prs{Y}$ is an \OU{} process driven by the following SDE,
\begin{equation*}
    dY_t = 0.25\ (0.005-Y_t)\ dt + 0.01\ dB^Y_t,\quad Y_0=0.005,
\end{equation*}
We have chosen the parameters $k = 0.25$, $\mu=0.005$, and $\sigma=0.01$. The process $B^Y=\prs{B^Y}$ is a standard Brownian motion satisfying $\EE\left[B_t^S B_t^Y\right]=\rho$, where the constant $\rho$ is left unspecified for now. Our simulation considers three different traders with asymmetric levels of information. The first investor operates naturally and follows the natural filtration~$\bF$ generated by $(B^S,B^Y)$. The second investor is partially informed and operates with an enlarged filtration incorporating the random variable $L = \bOne \{ c_1\leq Y_1\leq c_2 \}$, denoted as ${\overline{\bG}} = \bF\vee\sigma (\bOne{\{Y_1\in(c_1,c_2)\}})$. The third investor has precise information about the interest rate at the time horizon and operates with $\bG=\bF\vee\sigma(Y_1)$. Let's revisit the formulation of the optimization problem.
\begin{equation*}
    \bV^{\bH} = \sup_{\pi\in\cA(\bH)}\EE[\ln X_1^{\pi}],\quad X_0 = 1,\quad \bH\in\{\bF,{\overline{\bG}},\bG\}.
\end{equation*}
Based on the arguments developed in the previous sections, we ascertain that the optimal portfolios solving the aforementioned optimization problems are as follows:
\begin{align}
   \pi^{\bF}_t = \frac{0.05 - Y_t}{0.01^2},
   \quad
   \pi^{\overbar{\bG}}_t = \pi^{\bF}_t + \frac{\rho}{0.01} \alpha_t^L,
   \quad 
   \pi^{\bG}_t = \pi^{\bF}_t + \frac{\rho}{0.01} \alpha_t^{Y_1} ,
\end{align}
where $\alpha^L$ is defined in~\eqref{eq:alpha.interval.finite.ou} and $\alpha^{Y_1}$ in~\eqref{alpha.precise.OU}.
We simulate 100 trajectories of the assets $(D,S)$, and we denote them by $(D(k),S(k))$, with $k\in\{1,2\dots,100\}$.
We compute the optimal strategies for each information level $\pi_t^{\bH}(k)$ with \hbox{$k\in\{1,2\dots,100\}$} and $\bH\in\{\bF,\overline{\bG},\bG\}$,
as well as the corresponding utility gains $\ln X_1^{\pi^{\bH}}(k)$  computed according to~\eqref{X.exponential.solution}.

Table~\ref{tab:rho-corr} shows a summary of the results where the agents, playing with the filtrations $\bF$, $\overline{\bG}$ and $\bG$, are respectively denoted  as \emph{Natural}, \emph{Interval} and \emph{Precise}.

The row \emph{Gains} refers to the average gain on the 100 simulation, i.e.,
\begin{equation*}
    Gains(\bH) = \frac{1}{100}\sum_{k=1}^{100} \ln X_1^{\pi^{\bH}}(k).
\end{equation*}
In order to quantify the risk that the agents accept in each level of information, we compute,
\begin{equation*}
    \hat\pi_t(\bH) = \max_{k}\{\pi_t^{\bH}(k)\} - \min_k\{\pi_t^{\bH}(k)\},\quad 0\leq t \leq 1
\end{equation*}
and we give the quartiles and the range of $\hat\pi(\bH)$.
To show the importance of the correlation in our set-up, 
in Table~\ref{tab:rho-corr} we compare the optimal portfolios and the corresponding utilities for different values of $\rho \in\{ 0.75, 0.25, -0.75\}$.

The gains of the Natural agent do not depend on the value of  $\rho$, so it is given at the beginning as a reference value for the evaluation of the values of the information for the Interval and Precise agents.
For the Interval agent we assume that the interval $(c_1,c_2)$ is equal to $(0.7\,Y_1,1.3\,Y_1)$. 
\begin{table}[ht]
\begin{center}
\begin{tabular}{cccccccc}
    \hline
    \vspace{0.2 cm}
    {}&\multicolumn{3}{c}{$\rho=0.75$}&\multicolumn{2}{c}{$\rho=0.25$}&\multicolumn{2}{c}{$\rho=-0.75$}\\
    \hline
    \vspace{0.2 cm}
    {}&{Natural}&{Interval}&{Precise}&{Interval}&{Precise}&{Interval}&{Precise}\\
    \cline{1-8}
    $Gains$         &   0.25    &   0.53    &   4.71    &   0.32    &   0.66    &   0.55    &   4.53\\
    $Q_1(\pi^G)$    &   198   &   230   &   389   &   182   &   213   &   244   &   432\\
    $Q_2(\pi^G)$    &   298   &   310   &   478   &   315   &   338   &   304   &   379\\
    $Q_3(\pi^G)$    &   391   &   366   &   867   &   349   &   443   &   421   &   775\\
    $Range(\pi^G)$  &   442   &   453   &   2636  &   396   &   1040   &   472   &   2123\\
    \hline
\end{tabular}
\end{center}
\caption{Results for different values of correlation $\rho$ between $B^S$ and $B^Y$.}
\label{tab:rho-corr}
\end{table}

\begin{figure}[ht]\centering
    \includegraphics[width=1\textwidth]{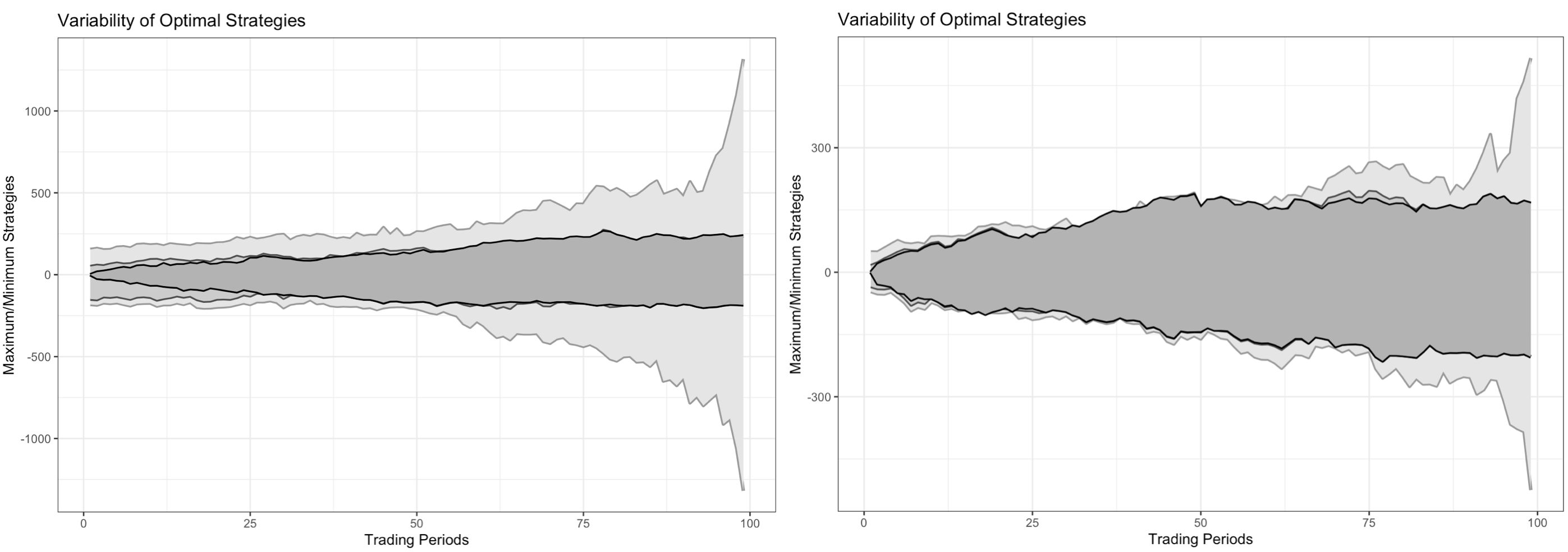}
     \caption{Plots for $\max_{k}\{\pi_t^{\bH}(k)\}$ and $\min_{k}\{\pi_t^{\bH}(k)\}$. The black is for the Natural trader, middle-gray for the Interval trader and light-gray for the Precise one.}\label{fig:1}
    \end{figure}

\begin{figure}[ht]\centering
    \includegraphics[width=1\textwidth]{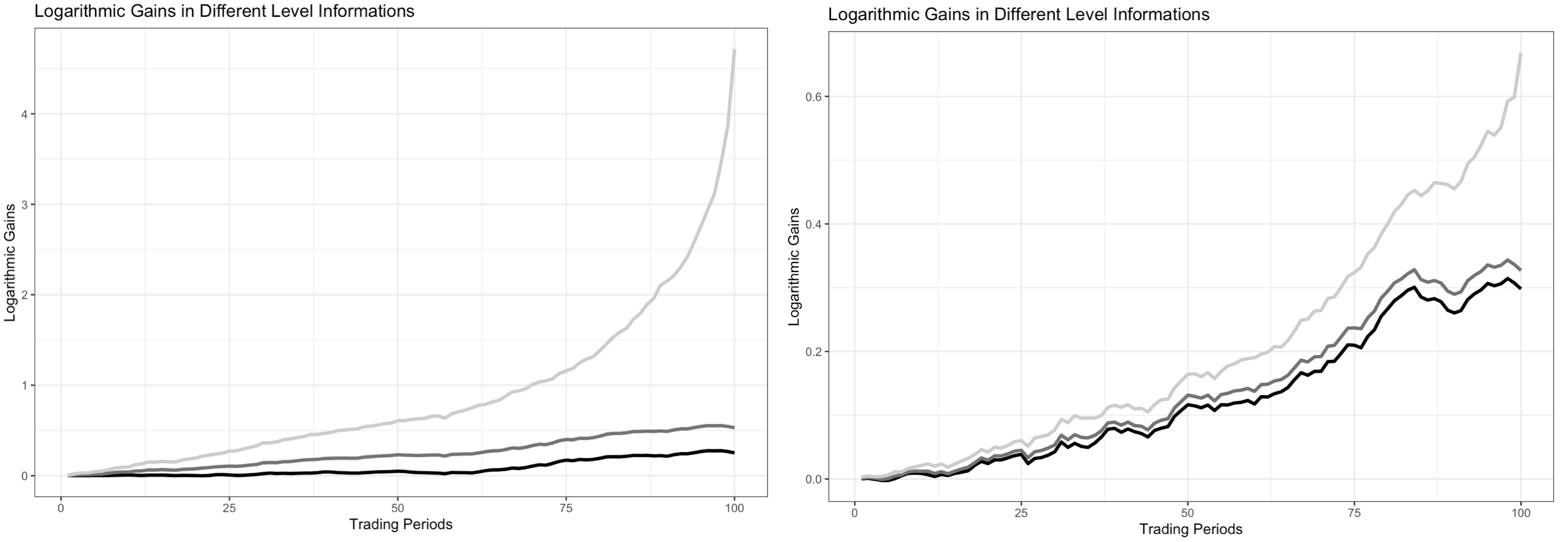}
   \caption{Comparison of the average of the logarithmic gains for $\rho = 0.75$ (left) and $\rho = 0.25$ (right). The black is for the Natural trader, middle-gray for the Interval trader and light-gray for the Precise one.}
   \label{fig:2}
\end{figure}

In Figure~\ref{fig:1}, we plot the values $\max_{k}\{\pi_t^{\bH}(k)\}$ and $\min_k \{\pi_t^{\bH}(k)\}$ and, in Figure~\ref{fig:2}, the average of the utility gain in each time, that is
\begin{equation*}
    \widehat{\ln X_t^{\pi^{\bH}}} = \frac{1}{100}\sum_{k=1}^{100} \ln X_t^{\pi^{\bH}}(k).
\end{equation*}
Note that $Gains(\bH)=\widehat{\ln X_1^{\pi^{\bH}}}$.
We only show the plots for $\rho = 0.75$ and $\rho = 0.25$, because the the ones for $\rho=-0.75$ look like the ones with $\rho=0.75$.


The simulations show that the logarithmic utility gains are greater if the Brownian motions $B^S$ and $B^Y$ are strongly correlated.
If the correlation is negative, the agent can achieve the same gains because investing against the risky asset $S$ is allowed.
It is also shown that more accurate information implies higher gains, at the cost of making a riskier investment.

\begin{table}[ht]
\begin{center}
\begin{tabular}{cccccc}
    \hline
    \vspace{0.2 cm}
    {}&{}&\multicolumn{2}{c}{$\rho=0.75$}\\
    \hline
    \vspace{0.2 cm}
    {}&{$v = 0.1$} &{$v=0.25$} &{$v=0.5$} &{$v=0.75$} &{$v = 2$}\\
    \cline{1-6}
    $Gains$ &   4.64    &   0.69    &   0.38    &   0.28    &   0.25\\
    $Q_1(\pi^G)$ &   410  &   325   &   224   &   178   &   195\\
    $Q_2(\pi^G)$ &   518  &   348   &   306   &   315   &   311\\
    $Q_3(\pi^G)$ &   620 &   366   &   363   &   357   &   378\\
    $Range(\pi^G)$&  969 &   423  &   415   &   410   &   454\\
    \hline
\end{tabular}
\end{center}
\caption{Results for different values of information for interval type.}
\label{tab:v-values}
\end{table}

In Table~\ref{tab:v-values}, we fix $\rho=0.75$ and study additional information of  interval type, for the filtration $\overline{\bG}$, where the interval is always of the following form
\begin{equation*}
(c_1,c_2) = ((1-v)Y_1,\ (1+v)Y_1),\quad 0<v<1.
\end{equation*}
Whenever $v$ approaches 0, the Interval agent behaves more similarly to the Precise one, as her information is more accurate, while as $v$ is near to 1, the Interval agent plays similarly to the Natural one since her information is less  accurate.
Again we omit the results for $\rho=-0.75$.

\begin{figure}[ht]\centering
    \includegraphics[width=0.75\textwidth]{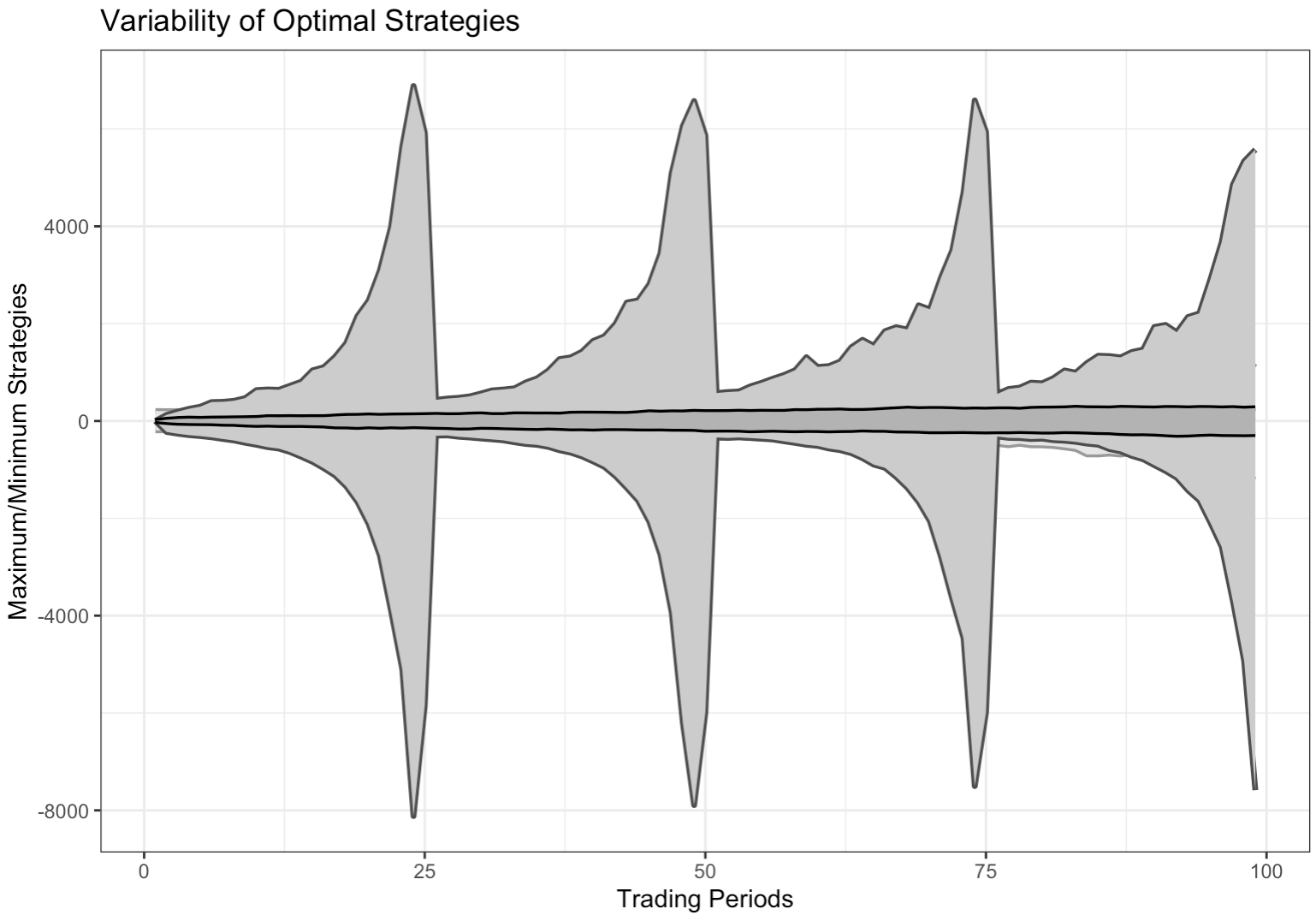}
     \caption{Plot for $\max_{k}\{\pi_t^{\bH}(k)\}$ and $\min_{k}\{\pi_t^{\bH}(k)\}$. The black is for the Natural trader, middle-gray for the Markov-modulated one and light-gray for the Precise one.}
    \label{fig:3}
\end{figure}

We complete the analysis by simulations with an example related to Section~\ref{sec:markov.modul}, in which the information is revealed to the agent only at certain times, by assuming that, in the market, there is a modulating process.
We introduce the Markov chain
\begin{equation*}
    \varepsilon_{t_\kappa} = \bOne_{ \{B^Y_{t_{\kappa+1}}-m_{t_{\kappa+1}}\leq  c \}}  ,\quad c\in\bR^+,\quad \kappa\in\{0,1,2,3\},
\end{equation*}
as in Example~\ref{Ex.min} and we use equation~\eqref{eq:drift.markov} and Proposition~\ref{Markov.infor.drift} to compute the information drift.
Figure~\ref{fig:3} shows the strategies adopted by the various agents, and, as expected, they show a pronounced variability close to each time epoch $t_\kappa$, $\kappa\in\{0,1,2,3\}$.

\begin{figure}[ht]\centering
\includegraphics[width=1\textwidth]{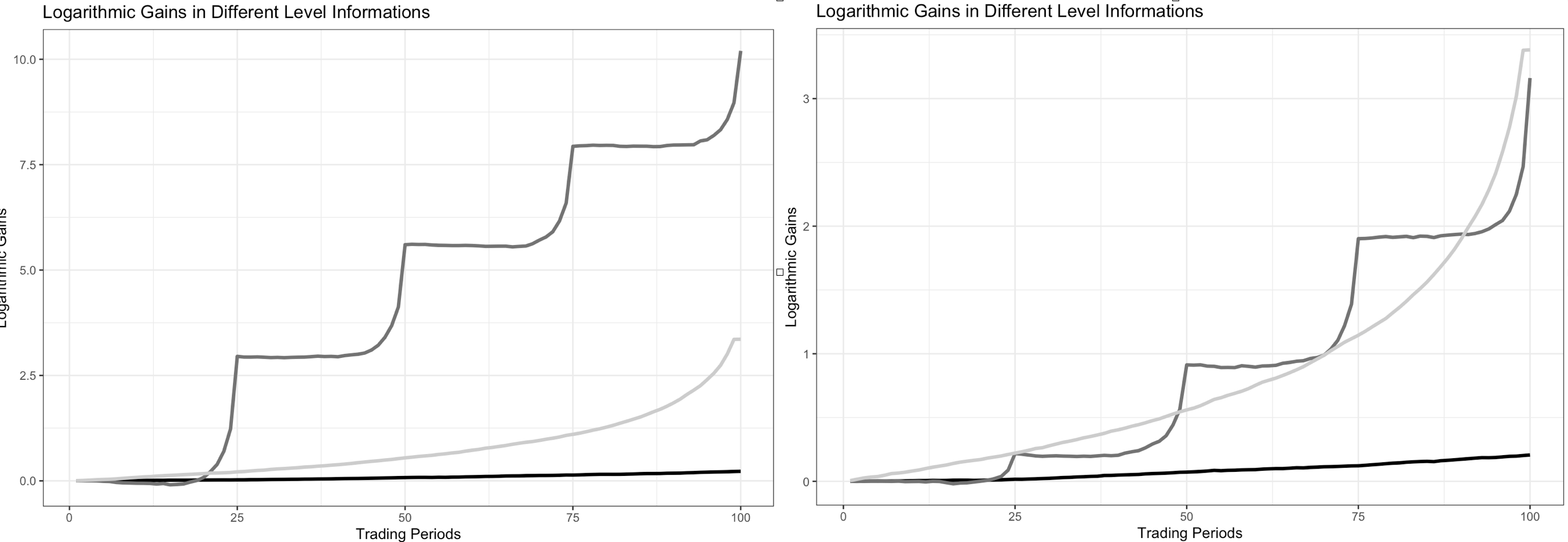}
 \caption{Comparison of the average of the logarithmic gains for $c = 0.25$ (left) and $c = 1.25$ (right). The black is for the Natural trader, middle-gray for the Markov-modulated one and light-gray for the Precise one. }
\label{fig:4}
\end{figure}

To compare with the previous simulations, in this analysis we keep considering the Honest and the Precise traders with $\rho=0.75$.
The constant $c$ allows to vary the accuracy of the information, indeed, a small value of $c$ means very accurate information whereas a larger value of $c$ implies less accurate information.
This appears evident in Figure~\ref{fig:4}, where we show the simulation of the gains for different values of $c$, in particular $0.25$ and $1.25$.

\section{Conclusions}\label{sec:conclusions}
Introducing anticipating information on the short rate interest in a portfolio optimization problem may be dealt with by techniques of enlargements of filtrations.
When the driving Brownian motion of the short rate process is correlated with the one of the risky asset, the informed agent may obtain unbounded gains depending on the accuracy of the anticipating information.
This paper has analyzed in detail the optimal portfolio problem when the short rate is a general affine diffusion, a class that allows for a more explicit analysis and that contains, in particular, the \OU{} process used in the \Vmod{} model. 
The kind of additional information analyzed is any that is modeled through the general class of $L^2(\cF_T^R,\PP)$ random variables.
Some explicit examples give, for instance, information modeled by characteristic functions of half-bounded and bounded intervals, as well as the case of the $n$-power of the terminal value of the driving Brownian motion.
The analysis is then extended to the case of a market that includes a Markov chain  modulating the evolution of the stochastic interest rate. 
For this case, and assuming anticipating information on the modulating chain, we show how to compute the additional logarithmic utility for an informed trader.

\section*{Acknowledgments}

This research was partially supported by the Spanish \emph{Ministerio de Economía y Competitividad} grant
PID2020-116694GB-I00.
The second author acknowledges financial support from an FPU Grant (FPU18/01101) of Spanish \emph{Ministerio de Universidades}.
\printbibliography
\appendix

\section{Appendix}\label{Appendix} 
\begin{proof}[\textit{Proof\,of\,Lemma\,\ref{lema2}}]
    \hypertarget{proof.lem.distr.R}
    By equation~\eqref{eq:solLineal}, we can express the value of $R_t$ in terms of its value at time $s$ in the following way,
    \begin{align*}
        \label{procNormal}
        R_t = & \Psi_{0,s}\Psi_{s,t}\Big[  R_0 + \int_0^s \Psi^{-1}_u a_2(u) du + \int_s^t \Psi^{-1}_u a_2(u) du                          \\
        & + \int_0^s \Psi^{-1}_u b_2(u) dB^R_u + \int_s^t \Psi^{-1}_u b_2(u) dB^R_u  \Big]                                          \\
        =     & \Psi_{s,t} R_s + \Psi_t\left[\int_s^t \Psi^{-1}_u a_2(u) du + \int_s^t \Psi^{-1}_u b_2(u) dB^R_u  \right] \numberthis \ .
    \end{align*}
    The result then follows by identifying the deterministic and stochastic part of formula~\eqref{procNormal}, and the latter gives the variance
    by applying the It\^o isometry.
\end{proof}
\begin{proof}[\textit{Proof\,of\,Proposition\,\ref{prop:semimart.repr.bG}}]
    \hypertarget{proof.prop.semimart.repr.bG}
    Since $B^R \text{ and }\,B^S$ have constant correlation $\rho$, we can write
    \begin{align}
        \label{corr.eq}
        B_t^S = \rho B_t^R + \sqrt[]{1-\rho^2}\, B_t\ , \quad 0\leq t \leq T\ ,
    \end{align}
    where $B = \prT{B}$ is an $\bF$-Brownian motion independent of $B^R$.
    Note that we can express $\cF_t = \sigma\left((B_s^R,B_s) : 0\leq s \leq t \right)$ and $(B^R,B)$ has the predictable representation property in $\bF$.
    Let $g\in Supp(G)$, then
    \begin{align*}
        \langle B,p^g\rangle_t^\bF & = \int_0^t p_s^g\alpha_s^g d\langle B,B^R\rangle_s^\bF = 0\ . 
    \end{align*}
    Then we conclude that $B$
    is a $\bG$-Brownian motion.
    Finally, from the independence of $B$
    and $W^R$ jointly with the fact that $G\in \cF^R_T$,
    we get the last part of the statement.

    The formula for the additional expected logarithmic utility  follows by applying \citet[Theorem 2.13]{AnkirchnerImkeller05}.
\end{proof}
\begin{proof}[\textit{Proof\,of\,Proposition\,\ref{drift.interval.infinite}}]
    \hypertarget{proof.drift.interval.infinite}
    By employing the explicit expression of the process $R$ given in \eqref{eq:solLineal} we can rewrite the information given by $A$ with respect to the Brownian motion $B^R$.
    Note that
    \begin{equation*}
        \{R_T\leq c\} = \Big\{  \int_0^T \Psi^{-1}_t b_2(t) dB^R_t \leq \tilde c \Big\}\ ,\ \text{with } \tilde c= \frac{c}{\Psi_T}-R_0-\int_0^T \Psi^{-1}_t a_2(t) dt\ .
    \end{equation*}
    It follows that the process $\alpha^A=\prTmenos{\alpha^A}$  satisfies the following relation
    \begin{equation*}
        dp_t^A = p_t^A \alpha^A_t dB^R_t\ ,\quad\text{with }p_t^a = \frac{\PP(A = a \vert \cF_t)}{\PP(A = a)}\ ,\quad a\in\{0,1\}\ .
    \end{equation*}
    The conditional probability mass function of $A$ computed for $\{A=1\}$ may be expressed in terms of Gaussian distribution $\Phi$ as
    \begin{equation*}
        \PP(A = 1 \vert \cF_t) = \Phi\left(\frac{\tilde c-\int_0^t \Psi^{-1}_u b_2(u) dB^R_u}{\sqrt{\int_t^T \Psi^{-2}_u b_2^2(u) du }}\right) \ .
    \end{equation*}
    By It\^o lemma, we can compute its derivative,
    \begin{align*}
        d\PP(A = 1 \vert \cF_t) & =  -\frac{\Psi^{-1}_t b_2(t) }{\sqrt{\int_t^T \Psi^{-2}_u b_2^2(u) du}} \Phi'\left(\frac{\tilde c-\int_0^t \Psi^{-1}_u b_2(u) dB^R_u }{\sqrt{\int_t^T \Psi^{-2}_u b_2^2(u) du}}\right)dB^R_t \\
        & = -\Psi_T \Psi^{-1}_t b_2(t) f_{R_T\vert R_t}(c\vert R_t) dB^R_t\ ,
    \end{align*}
    and finally,
    \begin{equation*}
        \frac{dp^1_t}{p^1_t} = -\Psi_T \Psi^{-1}_t b_2(t) \frac{ f_{R_T\vert R_t}(c\vert R_t)}{\PP( A=a \, \vert \cF_t)}dB^R_t\ ,
    \end{equation*}
    and we get the result. The case $\{A=0\}$ follows the same lines.
\end{proof}

\end{document}

%% file: commands.tex
\providecommand{\norm}[1]{\lVert#1\rVert}

\newcommand{\prT}[1]{(#1_t, \, 0 \leq{t} \leq{T})}

\newcommand{\prs}[1]{(#1_t, \, 0 \leq{t} \leq{1})}
\newcommand{\prTmenos}[1]{(#1_t, \, 0 \leq{t} <T)}
\newcommand\numberthis{\addtocounter{equation}{1}\tag{\theequation}}

\newcommand{\overbar}[1]{\mkern 1.5mu\overline{\mkern-1.5mu#1\mkern-1.5mu}\mkern 1.5mu}

\newcommand{\cA}{\mathcal{A}}
\newcommand{\cB}{\mathcal{B}}

\newcommand{\cF}{\mathcal{F}}
\newcommand{\cG}{\mathcal{G}}
\newcommand{\cH}{\mathcal{H}}
\newcommand{\cN}{\mathcal{N}}

\newcommand{\bF}{\mathbb{F}}
\newcommand{\bG}{\mathbb{G}}

\newcommand{\bH}{\mathbb{H}}
\newcommand{\bOne}{\mathbbm{1}}
\newcommand{\bN}{\mathbb{N}}

\newcommand{\bR}{\mathbb{R}}
\newcommand{\bV}{\mathbb{V}}
\newcommand{\PP}{\boldsymbol{P}}
\newcommand{\EE}{\boldsymbol{E}}
\newcommand{\VV}{\boldsymbol{V}}

\DeclareMathOperator{\Spp}{Supp}
 
\newcommand{\OU}{Ornstein-Uhlenbeck}
\newcommand{\Vmod}{Vasicek}
\newcommand{\defeq}{:=}
\newcommand{\disteq}{\sim}

%% file: main.bib
@ARTICLE {GibsonLhabitantTalay2010,
    author  = "Rajna Gibson and Francois-Serge Lhabitant and Denis Talay",
    title   = "Modeling the Term Structure of Interest Rates: A Review of the Literature",
    journal = "Foundations and Trends in Finance",
    year    = "2010",
    volume  = "5",
    number  = "1--2",
    pages   = "1--156",
    issn    = "1567-2395",
}

@ARTICLE {AmendingerImkellerSchweizer1998,
    author   = "Jürgen Amendinger and Peter Imkeller and Martin Schweizer",
    title    = "Additional logarithmic utility of an insider",
    journal  = "Stochastic Processes and their Applications",
    year     = "1998",
    volume   = "75",
    number   = "2",
    pages    = "263--286",
    issn     = "0304-4149"
}

@INCOLLECTION {Jacod1985,
    author    = "Jacod, Jean",
    title     = "Grossissement initial, hypoth{\'e}se {(H')} et th{\'e}or{\'e}me de {G}irsanov",
    booktitle = "Grossissements de filtrations: exemples et applications",
    publisher = "Springer",
    year      = "1985",
    pages     = "15--35",
    address = "Paris"
}

@ARTICLE {vasicek,
    author     = "Oldrich Vasicek",
    title      = "An equilibrium characterization of the term structure",
    journal    = "Journal of Financial Economics",
    year       = "1977",
    volume     = "5",
    number     = "2",
    pages      = "177--188",
    issn       = "0304-405X"
}

@INCOLLECTION {baudoin2003,
    author     = "Fabrice Baudoin",
    title      = "Modeling anticipations on financial markets",
    booktitle = "{Paris-Princeton} Lectures on Mathematical Finance 2002",
    publisher = "Springer",
    address = "Berlin, Heidelberg",
    year       = "2003",
    volume     = "1814",
    Pages = "43--94",
    issn       = "978-3-540-44859-4",
}

@BOOK {chesneyyorjeanblanc2009,
    author    = "Monique Jeanblanc and Marc Yor and Marc Chesney",
    title     = "Mathematical Methods for Financial Markets",
    publisher = "Springer London",
    year      = "2009",
    series    = "Springer Finance",
    isbn      = "9781852333768",
    address = "London"
}

@ARTICLE {karatzas1996,
    author    = "Igor Pikovsky and Ioannis Karatzas",
    title     = "Anticipative Portfolio Optimization",
    journal   = "Advances in Applied Probability",
    year      = "1996",
    volume    = "28",
    number    = "4",
    pages     = "1095--1122",
    issn      = "00018678"
}

@ARTICLE {Bauerle,
    author    = "Nicole Bauerle and Ulrich Rieder",
    title     = "Portfolio optimization with {M}arkov-modulated stock prices and interest rates",
    journal   = "IEEE Transactions on Automatic Control",
    year      = "2004",
    volume    = "15",
    number    = "3",
    pages     = "442--447",
    issn      = "0018-9286"
}

@ARTICLE {Amendinger2003,
    author  = "Amendinger, J{\"u}rgen  and Becherer, Dirk  and Schweizer, Martin",
    title   = "A monetary value for initial information in portfolio optimization",
    journal = "Finance and Stochastics",
    year    = "2003",
    volume  = "7",
    number  = "1",
    pages   = "29--46",
    issn    = "0949-2984"
}

@book{Aksamit2017,
  year = {2017},
  publisher = {Springer International Publishing},
  author = {Anna Aksamit and Monique Jeanblanc},
  title = {Enlargement of Filtration with Finance in View},
  address ={Berlin, Heidelberg}
}

@ARTICLE {AnkirchnerImkeller05,
    author     = "Stefan Ankirchner and Peter Imkeller",
    title      = "Finite utility on financial markets with asymmetric information and structure properties of the price dynamics",
    journal    = "Annales de l'Institut Henri Poincaré, Statistics",
    year       = "2005",
    volume     = "41",
    number     = "3",
    pages      = "479--503",
    issn       = "0246-0203"
}

@ARTICLE {Fontana18,
    author     = "Huy N Chau and Andrea Cosso and Claudio Fontana",
    title      = "The value of informational arbitrage",
    journal    = "Finance and Stochastic",
    year       = "2020",
    volume     = "24",
    pages      = "277--307"
}

@article{IMKELLER2000,
title = "Free lunch and arbitrage possibilities in a financial market model with an insider",
journal = "Stochastic Processes and their Applications",
volume = "92",
number = "1",
pages = "103--130",
year = "2001",
issn = "0304-4149",
author = "Peter Imkeller and Monique Pontier and Ferenc Weisz",
}

@article{Connor2010,
author = {Andreas Basse-O'Connor},
title = {Representation of {Gaussian} semimartingales with applications to the covariance function},
journal = {Stochastics},
volume = {82},
number = {4},
pages = {381--401},
year  = {2010},
publisher = {Taylor & Francis}
}

@article{DauriaSalmeron2020,
author = {Bernardo D'Auria and Jose A. Salmeron},
title = {Insider information and its relation with the arbitrage condition and the utility maximization problem},
journal = {Mathematical Biosciences and Engineering},
volume = {17},
number = {2},
pages = {998--1019},
year  = {2020}}

@Article{AntonelliRamponiScarlattijournal,
  author={Fabio Antonelli and Alessandro Ramponi and Sergio Scarlatti},
  title={{Option-based risk management of a bond portfolio under regime switching interest rates}},
  journal={Decisions in Economics and Finance},
  year=2013,
  volume={36},
  number={1},
  pages={47--70}
}

@article{Azevedo2014,
title = {Dynamic programming for a {Markov}-switching jump–diffusion},
journal = {Journal of Computational and Applied Mathematics},
volume = {267},
pages = {1--19},
year = {2014},
issn = {0377-0427},
author = {N. Azevedo and D. Pinheiro and G.-W. Weber}
}

@article{Rogers2020,
author = {Ernst, Philip A. and Rogers, L. C. G.},
title = {The Value of Insight},
journal = {Mathematics of Operations Research},
volume = {45},
number = {4},
pages = {1193--1209},
year = {2020}
}

@article{karatzas1991,
author = {Karatzas, Ioannis and Lehoczky, John P. and Shreve, Steven E. and Xu, Gan-Lin},
title = {Martingale and Duality Methods for Utility Maximization in an Incomplete Market},
journal = {SIAM Journal on Control and Optimization},
volume = {29},
number = {3},
pages = {702--730},
year = {1991}
}

@article{PhamQuenez2001,
author = {Huy\^en Pham and Marie-Claire Quenez},
title = {Optimal Portfolio in Partially Observed Stochastic Volatility Models},
volume = {11},
journal = {The Annals of Applied Probability},
number = {1},
publisher = {Institute of Mathematical Statistics},
pages = {210--238},
year = {2001}
}

@book{DiNunnoOksendalProske2009,
   title =     {Malliavin Calculus for {Lévy} Processes with Applications to Finance},
   author =    {Giulia {Di Nunno} and Bernt Øksendal and Frank Proske},
   publisher = {Springer},
   isbn =      {354078571X},
   year =      {2009},
   series =    {Universitext},
   edition =   {1st ed. 2009. Corr. 2nd printing},
   volume =    {},
   address = {Berlin-Heidelberg}
}

@book{VectorMeasures77,
AUTHOR = {Diestel, Joseph and Uhl, Jerry Jerry},
TITLE = {Vector measures},
PUBLISHER = {American Mathematical Society},
ADDRESS = {Providence, R.I.},
YEAR = {1977},
PAGES = {xiii+322}
}

@article{Kakihara11,
author = {Y. Kakihara},
title = {Radon-{N}ikodým Derivatives of {H}ilbert Space Valued Measures},
journal = {Journal of Statistical Theory and Practice},
volume = {5},
number = {3},
pages = {453--473},
year  = {2011},
publisher = {Taylor & Francis}
}

@article{Imkeller2003,
author = {Imkeller, Peter},
title = {Malliavin's Calculus in Insider Models: Additional Utility and Free Lunches},
journal = {Mathematical Finance},
volume = {13},
number = {1},
pages = {153--169},
year = {2003}
}

@InProceedings{chaleyatJeulin,
author="Chaleyat-Maurel, Mireille
and Jeulin, Thierry",
editor="Jeulin, Th.
and Yor, M.",
title="Grossissement gaussien de la filtration {B}rownienne",
booktitle="Grossissements de filtrations: exemples et applications",
year="1985",
publisher="Springer Berlin Heidelberg",
address="Berlin, Heidelberg",
pages="59--109",
isbn="978-3-540-39339-9"
}

@article{JAPoksendal_2007, 
title={Optimal Smooth Portfolio Selection for an Insider},
volume={44}, 
number={3}, 
journal={Journal of Applied Probability}, 
publisher={Cambridge University Press}, 
author={Hu, Yaozhong and Øksendal, Bernt}, 
year={2007}, 
pages={742--752}}

@article{JAPdeelstra_grasselli_koehl_2000, 
title={Optimal investment strategies in a {CIR} framework},
volume={37}, 
number={4}, 
journal={Journal of Applied Probability}, 
publisher={Cambridge University Press}, 
author={Deelstra, Griselda and Grasselli, Martino and Koehl, Pierre-François}, 
year={2000},
pages={936--946}}

@article{GhoshGoswamiKumar2009,
title = {Portfolio Optimization in a Semi-{Markov} Modulated Market},
author = {Ghosh, Mrinal K. and Goswami, Anindya and Kumar, Suresh K.},
journal = {Applied Mathematics and Optimization},
issn = {0095-4616},
number = 2,
volume = 60,
place = {United States},
year = {2009},
pages = {275--296}
}

@book{OksendalSDE,
   title =     {Stochastic Differential Equations},
   author =    {Bernt Øksendal},
   publisher = {Springer},
   isbn =      {978-3-540-04758-2},
   year =      {2003},
   series =    {Universitext},
   edition =   {6},
   volume =    {},
   address = {Berlin, Heidelberg}
}

@article{SotomayorCadenillas2009,
  author={Luz Rocío Sotomayor and Abel Cadenillas},
  title={Explicit Solutions Of Consumption‐Investment Problems In Financial Markets With Regime Switching},
  journal={Mathematical Finance},
  year=2009,
  volume={19},
  number={2},
  pages={251--279}
}

@article{DAuriaSalmeronANOR,
  author={Bernardo D'Auria and Jose A. Salmeron},
  title={Anticipative information in a {Brownian-Poisson} market.},
  journal={Annals of Operations Research},
  year=2024,
  volume={336},
  number={1},
  pages={1289--1314}
}

@book{Privault2012interestrate,
author = {Privault, Nicolas},
title = {An Elementary Introduction to Stochastic Interest Rate Modeling},
publisher = {World Scientific},
year = {2012},
address = {Singapore},
edition   = {2 ed.}
}

@book{Filipovic09,
	year = {2009},
	publisher = {Springer Finance},
	author = {Damir Filipovi{\'{c}}},
	title = {Term-Structure Models},
    address = {Berlin}
}

@book{Alfonsi2015,
  year = {2015},
  publisher = {Springer International Publishing},
  author = {Aur{\'{e}}lien Alfonsi},
  title = {Affine Diffusions and Related Processes: Simulation,  Theory and Applications}
}

@article{Halconruy23DEF,
author = {Halconruy, Hélène},
year = {2023},
month = {09},
title = {The insider trading problem in a jump-binomial model},
journal = {Decisions in Economics and Finance},
volume ={46},
pages ={379--413}}

@Article{Zhang23DEF,
  author={Yumo Zhang},
  title={{Utility maximization in a stochastic affine interest rate and CIR risk premium framework: a BSDE approach}},
  journal={Decisions in Economics and Finance},
  year=2023,
  volume={46},
  number={1},
  pages={97--128}}

@article{Ferrari18, 
title={On an optimal extraction problem with regime switching}, 
volume={50}, 
number={3}, 
journal={Advances in Applied Probability}, 
publisher={Cambridge University Press}, 
author={Ferrari, Giorgio and Yang, Shuzhen}, 
year={2018}, 
pages={671--705}}

@article{Ferrari17,
  title={On the singular control of exchange rates},
  author={Giorgio Ferrari and Tiziano Vargiolu},
  journal={Annals of Operations Research},
  year={2017},
  volume={292},
  pages={795--832}
}

@article{Amendinger00,
title = {Martingale representation theorems for initially enlarged filtrations},
journal = {Stochastic Processes and their Applications},
volume = {89},
number = {1},
pages = {101--116},
year = {2000},
issn = {0304-4149},
author = {Jürgen Amendinger},
}

@article{AnkirchnerImkeller06,
 ISSN = {00911798},
 author = {Stefan Ankirchner and Steffen Dereich and Peter Imkeller},
 journal = {The Annals of Probability},
 number = {2},
 pages = {743--778},
 publisher = {Institute of Mathematical Statistics},
 title = {The {Shannon} Information of Filtrations and the Additional Logarithmic Utility of Insiders},
 volume = {34},
 year = {2006}
}

@article{Biagini2005,
  year = {2005},
  month = jul,
  publisher = {Springer Science and Business Media {LLC}},
  volume = {52},
  number = {2},
  pages = {167--181},
  author = {Francesca Biagini and Bernt {\O}ksendal},
  title = {A General Stochastic Calculus Approach to Insider Trading},
  journal = {Applied Mathematics and Optimization}
}
